\documentclass[11pt]{article}
\pdfoutput=1
\usepackage{amsmath,epsf,amssymb,latexsym,array,color}
\usepackage{graphicx,wasysym}
\usepackage{cancel}

\def\be{\begin{equation}}
\def\ee{\end{equation}}
\def\ba{\begin{array}}
\def\ea{\end{array}}
\newcommand{\beq}{\begin{equation}}
\newcommand{\eeq}[1]{\label{#1}\end{equation}}
\newcommand{\bea}{\begin{eqnarray}}
\newcommand{\eea}[1]{\label{#1}\end{eqnarray}}

\usepackage{ifpdf}
\ifx\pdfoutput\undefined
   \pdffalse
\else
   \pdfoutput=1
   \pdftrue
  \usepackage[pdftex]{hyperref}
\begin{document}
%\title{Constraining Liberated Supergravity}
%\author{Hun Jang}
%\email{hun.jang@nyu.edu}
%\author{Massimo Porrati}
%\email{massimo.porrati@nyu.edu, hun.jang@nyu.edu}
%\affiliation{Center for Cosmology and Particle Physics\\ Department of Physics, New York University \\
%726 Broadway, New York NY 10003, USA}

%\maketitle

\begin{titlepage}

\hskip 1.5cm

\begin{center}
{\huge \bf{A Realization of Slow Roll Inflation and the MSSM in Supergravity Theories with New Fayet-Iliopoulos Terms}}
\vskip 0.8cm  
{\bf \large Hun Jang\footnote{hun.jang@nyu.edu} and Massimo Porrati\footnote{massimo.porrati@nyu.edu}}  
\vskip 0.75cm

{\em Center for Cosmology and Particle Physics\\
	Department of Physics, New York University \\
	726 Broadway, New York, NY 10003, USA}
	
	\vspace{12pt}

\end{center}

\begin{abstract}
A new supergravity D-term, not associated to gauged R-symmetry, was recently discovered and used to construct new supergravity 
models. In this paper we use a generalization of the new D-term that we used in previous works, to construct a supergravity model
of slow-roll inflation with the observable sector of the minimal supersymmetric standard model. Supersymmetry is broken at a high 
scale in the hidden sector and communicated to the observable sector by gravity mediation. The new D-term contains free parameters 
that can give large masses to scalar superpartners of quarks and leptons and to the higgsinos while holding the masses of 
observed particles fixed. Gauginos receive a mass from a non-canonical kinetic term for the vector supermultiplets.
We also present a simple argument proving in full generality that the cutoff $\Lambda$ of effective theories containing new D-terms 
can never exceed the supersymmetry breaking scale. In our theory, the relation between D-term and the Hubble constant 
during inflation also implies the universal relation $\Lambda \lesssim \sqrt{H M_{Pl}}$.

\end{abstract}

\vskip 1 cm
\vspace{24pt}
\end{titlepage}
\tableofcontents

\section{Introduction}
The effective field theory describing the low-energy dynamics of superstrings is described by a supergravity theory. If supersymmetry is 
broken at an energy scale comparable to the string scale $M_S\sim (\alpha')^{-1/2}$ it can be realized nonlinearly~\cite{nonlin} while if $M_S\ll (\alpha')^{-1/2}$ the constraints following from linearly realized supersymmetry restrict the effective action to the general  form 
found long ago in~\cite{cfgvnv}. This makes the construction of specific 
models of inflation in superstring theory quite challenging. In fact, no ``standard,'' canonical model of supergravity inflation exists 
today, even without requiring a string theory origin for it. 
When the supersymmetry breaking scale is higher than the Hubble scale during slow-roll inflation, $H$, 
nonlinear realization of supersymmetry give additional flexibility in constructing scalar potentials for the inflaton. In fact, they may be 
unnecessarily generic, since they do not require to have the same number of bosonic and fermionic degrees of freedom, which is
instead a robust prediction of any superstring theory. 

A novel method for enlarging the space of supergravity effective theories while automatically ensuring that each bosonic degree 
of freedom has fermionic  partner does exist. The method uses linear realizations of supersymmetry so all interactions can be written
in a manifestly supersymmetric form in terms of superfield. The difference with the general construction of~\cite{cfgvnv}  
is that the new interactions
become singular when supersymmetry is unbroken, because they contain {\em inverse} powers of some auxiliary fields.  Among these
new terms we will be particularly interested in the ``new Fayet-Iliopoulos'' (FI) terms written in a K\"ahler invariant, field-dependent 
form by Aldabergenov, Ketov, and Knoops (AKK) \cite{AKK}. This is a generalization of  the new FI terms introduced by Antoniadis, 
Chatrabhuti, Isono and Knoops (ACIK)~\cite{acik} as well as of the first new D-term to be discovered, constructed in~\cite{cftp}.
We used ACIK terms in our previous paper~\cite{jpd} to construct a toy model of slow-roll inflation with semi-realistic particle
spectrum based on the 
Kachru-Kallosh-Linde-Trivedi (KKLT) model \cite{KKLT} of superstring inflation. Its effective field theory description is a 
supergravity model with a no-scale K\"ahler potential for its volume modulus 
and with a superpotential that differs from its constant no-scale form because of non-perturbative 
corrections. 
The KKLT superpotential produces a supersymmetric Anti-de-Sitter (AdS) vacuum, which we lifted using the ACIK FI term. 
 In the model described in~\cite{jpd} supersymmetry was spontaneously broken in a hidden sector at a very high but still sub-Planckian scale 
$M_{pl} \gg M_S \gg 10^{-15}M_{pl}$. 
We employed gravity mediation to communicate the SUSY breaking to the observable sector, where supersymmetry breaking 
manifests itself through the existence of explicit soft 
SUSY breaking terms, characterized by an energy scale $M_{observable} \ll M_S$. Ref.~\cite{ACK} instead used AKK D-terms, a no-scale K\"ahler potential and a linear superpotential to obtain a slow-roll model of inflation.

The model we proposed in~\cite{jpd} had a major phenomenological weakness, originating from the fact that the scalars of the observable sector possess a universal contribution to the square mass $m^2$,
\beq
m^2= -{9\over 4} H^2 +... ,
\eeq{m1}
where $...$ denote small model-dependent corrections. The large term $-9H^2/4$ saturates the Breitenlohner-Freedman 
bound~\cite{bf} on the supersymmetric Anti de Sitter vacuum, where the scalar potential takes the value $V_{AdS}=-3H^2/8\pi G$.
When the vacuum energy is lifted up by a positive constant due to a nonzero D-term contribution, $V\rightarrow V + D^2/2 $,
the scalars become tachyons. We corrected for this problem by adding a large soft 
supersymmetry breaking terms in the superpotential, which made the masses of all scalar fields in the theory non-tachyonic but also
gave unacceptably large masses to the fermions in the observable sector. 
\begin{quote}
The first aim of this paper is to propose a modification of the ACIK-FI term that makes all scalars non-tachyonic, without requiring the 
introduction of large soft-supersymmetry breaking terms. 
\end{quote}
We will achieve a realistic spectrum of bosons and fermions in the observable sector by using the AKK FI term and by introducing 
a non-canonical kinetic term for some vector supermultiplets of the observable sector.

The new FI terms come together with a host of multi-fermion non-renormalizable terms, that are cumbersome to write and hard to 
study. The analysis carried out in~\cite{jpd} suffered from three weaknesses. The first one was that it was not systematic, because it 
examined only some non-renormalizable terms that had the structures necessary to give the strongest constraints on the UV cutoff of
the effective theory. The second one is that some of the potentially dangerous terms could vanish due to Fierz identities and other 
properties of multi-fermion terms. The third and most significant one is that the gauginos were not canonically normalized, so that 
spurious powers of the gauge coupling constant associated to the FI vector multiplet appeared in the formulas for the UV cutoff.
\begin{quote}
The second aim of this paper is to find the UV cutoff of the theories with new FI terms. We will find that ``three wrongs make a right''
and that these theories {\em can} have a cutoff $\Lambda \gg H$ and so they can be reliable effective field theories of inflation.
\end{quote}
The downside of our new analysis is that we will show in full generality that the cutoff of our theory cannot be parametrically larger
than $M_S$. 

This paper is organized as follows. In section~\ref{fi} we briefly review the construction of the AKK-FI term \cite{AKK} using the
superconformal tensor calculus language used in~\cite{jpd} and we apply it to lift the AdS minimum of the KKLT model.
In section~\ref{cutoff} we give a simple, general and model-independent argument showing that the UV cutoff $\Lambda$ in all theories with a new FI term obeys
\beq
\Lambda \lesssim M_S, \qquad \Lambda \lesssim \sqrt{H M_{Pl}}, \qquad M_{Pl}\equiv 1/\sqrt{8\pi G}.
\eeq{m2}
Section~\ref{infla} reviews the construction of a slow-roll ``Starobinsky"-type inflationary potential given in~\cite{jpd}, which uses the new FI term. The section also briefly discusses the hidden-sector dynamics due to the potential. 
In section~\ref{mass} we construct a superpotential that communicates the supersymmetry breaking due to the D term to the 
observable sector and produces a realistic spectrum of observable particles. In particular, we show how to keep standard models 
fermions light while making their scalar superpartners heavy, how to give mass to all the gauginos, how to keep the physical Higgs
scalar light while making the higgsino heavy, and how to ensure that a light Higgs field mass during inflation does not spoil the 
properties of single-field single field inflation. The AKK  generalization of the FI term, that replaces the FI constant term with a 
function of the
scalar fields in chiral multiplets, is used to make all scalars masses non-tachyonic in the post-inflation vacuum while a non-minimal
kinetic term for the gauge fields of the observable sector is used to give masses to the gauginos.  
An appendix summarizes  fermion masses formulas and their derivation in superconformal tensor calculus. Sections 2, 4, 5 and 
Appendix A use results derived in Ch. 12 and Appendix C of the doctoral dissertation of H.J. \cite{HJ_Dr_thesis}.

\section{New Fayet-Iliopoulos (FI) terms and KKLT}\label{fi}

In this section we construct a superconformal action of $\mathcal{N}=1$ supergravity equipped with generalized K\"{a}hler-invariant, field-dependent Fayet-Iliopoulos (FI) terms \cite{AKK} using superconformal tensor calculus \cite{STC,fvp}. To do this, we add to the standard $\mathcal{N}=1$ supergravity action a generalization of the new FI terms studied in Ref. \cite{jp2}. We write the superconformal action as 
\begin{eqnarray}
\mathcal{L} &=& -3[S_0\Bar{S}_0e^{-K(Z^A,\bar{Z}^{\bar{A}})/3}]_D + [S_0^3W(Z^A)]_F + \frac{1}{2}[
\mathcal{W}_{\alpha}(V)\mathcal{W}^{\alpha}(V)]_F + c.c. + \mathcal{L}_{new~FI} \label{newFI2}
\end{eqnarray}
where $S_0$ is the conformal compensator with Weyl/chiral weights (1,1); $Z^A$ and $V$ are chiral matter and vector multiplets with weights $(0,0)$; $K(Z,\bar{Z})$ is a K\"{a}hler potential gauged by a vector multiplet $V$, $W(Z)$ is a superpotential, and
$\mathcal{W}_{\alpha}(V)$ is the field strength of the vector multiplet $V$. 

Then, we decompose the matter multiplets $Z^A$'s into hidden and observable sectors, denoting with $T$ the volume modulus multiplet, and $Z^i$ the observable sector matter multiplets. We then consider a superpotential of the form
\begin{eqnarray}
 W(T) \equiv W^h(T)+W^o(Z^i), 
\end{eqnarray}
where
\begin{eqnarray}
 W^h(T) &\equiv& W_0 + Ae^{-aT}, \\
 W^o(Z^i) &\equiv& W_{\textrm{MSSM}} + ......,\\
 W_{MSSM} &\equiv&  -Y_u \hat{U}_R \hat{H}_u \cdot \hat{Q} + Y_d \hat{D}_R \hat{H}_d \cdot \hat{Q} + Y_e \hat{E}_R \hat{H}_u \cdot \hat{L}  + \mu \hat{H}_u \cdot \hat{H}_d \nonumber\\
 &=&  -Y_u \tilde{u}_R (H^+_u \tilde{d}_L -H^0_u\tilde{u}_L) + Y_d \tilde{d}_R (H^0_d \tilde{d}_L -H^-_d\tilde{u}_L)\nonumber\\
 && + Y_e \tilde{e}_R (H^+_u \tilde{\nu}_L -H^0_u\tilde{e}^-_L)  + \mu (H^+_uH^-_d-H^0_uH^0_d),
\end{eqnarray} 
where $W_0,A,a,Y_u,Y_d,Y_e,\mu$ are constants; $\hat{A} \cdot \hat{B} \equiv \epsilon_{ab}A^aB^b $ is the product between $SU(2)_L$ doublets in which $\epsilon_{12}=1=-\epsilon_{21}$ and $a,b$ are $SU(2)_L$ indices; $\tilde{u},\tilde{d},\tilde{e},\tilde{\nu}$ are the scalar component fields of the superfield $SU(2)_L$ doublets $\hat{Q},\hat{L}$. They are scalar superpartners of
 the SM quarks and leptons.

In our setup, the hidden sector superpotential $W^h$ has the form of a string theory superpotential with nonperturbative corrections, 
which are obtained by either Euclidean D3 branes in type IIB compactifications or gaugino condensation due to D7 branes \cite{KKLT}.
 The observable sector superpotential is the one used in minimal supersymmetric extensions of the standard model 
(MSSM)\footnote{We follow the notations used in Ref. \cite{GSR}.} plus ellipsis that stand for beyond the Standard Model corrections which we won't need in this paper. Its supermultiplets contain quarks, leptons, the Higgs fields and 
their supersymmetric partners. The K\"{a}hler potential of the volume modulus $T$ is the same as in KKLT string
 background \cite{KKLT} so it is given by
\begin{eqnarray}
 K = -3\ln[T+\bar{T}-\Phi(Z^i,\bar{Z}^{\bar{i}})/3],
\end{eqnarray}
where $\Phi(Z^i,\bar{Z}^{\bar{i}})$ is a real function of the observable-sector matter multiplets $Z^i$'s. In terms of the real function $\Phi$  
\begin{eqnarray}
 \Phi(Z^i,\bar{Z}^{\bar{i}}) = \delta_{i\bar{j}}Z^{i}\bar{Z}^{\bar{j}},
\end{eqnarray}
we have $\Phi_i\Phi^{i\bar{j}}\Phi_{\bar{j}}=\Phi$ where $\Phi_i \equiv \partial \Phi / \partial z^i$ and $\Phi_{i\bar{j}}\Phi^{l\bar{j}}=\delta_{i}^{l}$.

The next step is to determine which type of new FI terms we should be using. In this work, we employ the K\"{a}hler-invariant ``field-dependent'' Fayet-Iliopoulos (FI) terms proposed by Aldabergenov, Ketov, and Knoops (AKK) \cite{AKK}, which is a generalization of the ACIK-FI term \cite{acik}. We refer to this FI term as AKK-FI term to distinguish it from many other FI terms. The AKK-FI term 
that we will use is
\begin{eqnarray}
\mathcal{L}_{1} \equiv -  \left[(S_0\bar{S}_0e^{-K/3})^{-3}\frac{(\mathcal{W}_{\alpha}(V)\mathcal{W}^{\alpha}(V))(\bar{\mathcal{W}}_{\dot{\alpha}}(V)\bar{\mathcal{W}}^{\dot{\alpha}}(V))}{T(\bar{w}^2)\bar{T}(w^2)}(V)_D (\xi_1+U_1(\Phi, \bar \Phi, H, \bar H, V))\right]_D,\label{newFIterm} 
\end{eqnarray}
where 
$w^2 \equiv \frac{\mathcal{W}_{\alpha}(V)\mathcal{W}^{\alpha}(V)}{(S_0\bar{S}_0e^{-K(Z,\bar{Z})})^{2}}$ and $\bar{w}^2 \equiv \frac{\bar{\mathcal{W}}_{\dot{\alpha}}(V)\bar{\mathcal{W}}^{\dot{\alpha}}(V)}{(S_0\bar{S}_0e^{-K(Z,\bar{Z})})^{2}}$ are composite 
multiplets, $T(X), \bar{T}(X)$ are chiral projectors, and $(V)_D$ is a real multiplet, whose lowest component is the auxiliary 
field $D$ of the vector multiplet $V$; $\xi$ is a non-vanishing constant, and $U$ is a function of the chiral multiplets.
The solution for the auxiliary field $D$ for the vector multiplet is 
\begin{eqnarray}
 D /g^2 = \xi + U \equiv \mathcal{U}. \label{d-term}
\end{eqnarray}
In this work all fields are neutral under the $U(1)$ gauge symmetry gauged by the vector inside the $V$ supermultiplet so that 
no additional terms appear in the D-term equation~\eqref{d-term}.
Then, the $D$-term scalar potential is given by
\begin{eqnarray}
 V_D = \frac{1}{2g^2}D^2 = \frac{1}{2}g^2(\xi + U)^2 = \frac{1}{2}g^2\xi^2 + g^2\xi U + \frac{1}{2}g^2U^2.
\end{eqnarray}

Next we decompose the generic function $U$ into two pieces as 
\begin{eqnarray}
 U \equiv U^h + U^o \implies D/g^2 = \mathcal{U} = \xi+U=\xi + U^h + U^o,
\end{eqnarray}
$U^h$ depends on both hidden and observable sector fields and we will use it to give large masses to unobserved scalars. We 
will find it convenient to consider it as part of the hidden sector potential. 
$U^o$ depends on observable sector fields; we will use it to give
masses and expectation values to the Higgs fields and we will consider it as part of the observable sector potential.

The potential also depends on additional D-term potentials due to independent vector multiplets corresponding to the gauge symmetries of the theory. So, for different vector multiplets $V_A$, we have
\begin{eqnarray}
 V_D' \equiv \sum_A V_D^A = \sum_A \frac{g_A^2}{2}(k^I_A(z) G_I+c.c.)^2, 
\end{eqnarray} 
the gauge group contains of course  $U(1)_Y\times SU(2)_L\times SU(3)_C$.

Since the total supergravity scalar potential is given by the sum of F- and D-term potentials ($V_D$ and $V_F = e^G(G_AG^{A\bar{B}}G_{\bar{B}}-3)$), we find 
\begin{eqnarray}
 V =V_D + V_F = \Big( \frac{g^2}{2}(\xi+U^h+U^o)^2 + V_D' \Big) + V_F \equiv  V_h + V_{soft}.
\end{eqnarray}
Here we decomposed the scalar potential into hidden-sector and soft contributions according to the separation of the superpotential $W = W^h + W^o$, and thus the most dominant terms with $\sim |W^h|^2$ are combined into the hidden sector potential $V^h$:
\begin{eqnarray}
V_h &\equiv&\frac{g^2}{2}(\xi+U^h)^2 - \frac{W^h_T\bar{W}^h+\bar{W}^h_{\bar{T}}W^h}{X^{2}} + \frac{|W^h_T|^2}{3X^{2}}\Big(X+\frac{1}{3}\Phi_i\Phi^{i\bar{j}}\Phi_{\bar{j}}\Big) \nonumber\\
&& + \frac{1}{3}\frac{1}{X^{2}}
[W^h_T\Phi_{i}\Phi^{i\bar{j}}\bar{W}^h_{\bar{j}}
+\bar{W}^h_{\bar{T}}W^h_{i}\Phi^{i\bar{j}}\Phi_{\bar{j}}]+\frac{1}{X^{2}}W^h_i\Phi^{i\bar{j}}\bar{W}^h_{\bar{j}} ,
\eea{v-hidden}
while the rest of the potential is
\bea
V_{soft} &\equiv& g^2(\xi+U^h)(U^o) + \frac{g^2}{2}(U^o)^2 + V_D'-\frac{1}{X^{2}}[W^o\bar{W}^h_{\bar{T}}+\bar{W}^oW^h_{T}] \nonumber\\&&
+ \frac{1}{3}\frac{1}{X^{2}}
[W^h_T\Phi_{i}\Phi^{i\bar{j}}\bar{W}^o_{\bar{j}}
+\bar{W}^h_{\bar{T}}W^o_{i}\Phi^{i\bar{j}}\Phi_{\bar{j}}]
+\frac{1}{X^{2}}W^o_i\Phi^{i\bar{j}}\bar{W}^o_{\bar{j}}.
\eea{v-obs}
Here $X \equiv T+\bar{T}-\Phi(z,\bar{z})/3$ and $W_I \equiv \partial W/\partial z^I$ for $I=T,i$. Inserting the superpotentials into the above equation we obtain, similarly to Ref. \cite{jp2}
\begin{eqnarray}
V_h &=& \Big( \frac{1}{2}g^2\xi^2 + g^2\xi U^h  + \frac{1}{2}g^2{U^h}^2 \Big)  - \frac{1}{X^2}\bigg(-2aA^2e^{-a(X+\Phi/3)}+2acA^2 e^{-a(X+\Phi/3)/2} \cos(a\textrm{Im}T)\bigg)
\nonumber\\&&+ \frac{1}{3X^{2}}\Big(X+\frac{1}{3}\Phi_i\Phi^{i\bar{j}}\Phi_{\bar{j}}\Big) a^2A^2 e^{-a(X+\Phi/3)},\\
V_{soft} &\equiv& g^2(\xi+U^h)(U^o) + \frac{g^2}{2}(U^o)^2 +  \sum_A \frac{g_A^2}{2}(k^I_A(V_A) G_I+c.c.)^2 \nonumber\\
&&+
\frac{aAe^{-a(X+\Phi/3)/2}}{X^{2}}[W^oe^{ia\textrm{Im}T}+\bar{W}^oe^{-ia\textrm{Im}T}]
+\frac{1}{9}\frac{a^2A^2e^{-a(X+\Phi/3)}}{X^{2}}\Phi_{i}\Phi^{i\bar{j}}\Phi_{\bar{j}} \nonumber \\ &&
- \frac{1}{3}\frac{aAe^{-a(X+\Phi/3)/2}}{X^{2}}
[e^{-ia\textrm{Im}T}\Phi_{i}\Phi^{i\bar{j}}\bar{W}^o_{\bar{j}}
+e^{ia\textrm{Im}T}W^o_{i}\Phi^{i\bar{j}}\Phi_{\bar{j}}] 
+\frac{1}{X^{2}}W^o_i\Phi^{i\bar{j}}\bar{W}^o_{\bar{j}},
\end{eqnarray}
In the previous formulas we defined $W_0 \equiv -cA$, where $c$ is a constant.

\section{The UV Cutoff Due to new FI terms}\label{cutoff}

Once the  fields in the vector multiplet $V$ are canonically normalized, their D-terms appear in the Lagrangian   density as follows
\beq
L= {1\over 2} D^2 - \Xi D + \sum_{n\geq 0, m\geq 0} D^{-n} M_{Pl}^{-m} O^{2n+m +4} .
\eeq{m3}
Here $\Xi$ has scaling dimension two and we kept the 
dependence on $M_{Pl}$ explicit. The non-renormalizable operators $O^{2n+m +4}$ have scaling dimension 
$\Delta=2n+m +4$; they contain at lest two fermions and were given explicitly in~\cite{jp2}. Eq.~\eqref{m3} is an obvious consequence
of dimensional analysis and requires no knowledge of the explicit form of the operators $O^{2n +m +4}$. 
The key point here is that the powers of $M_{Pl}$ appearing in the non-renormalizable terms are always non-positive. This is because
there exist only two types of non-renormalizable interactions. The first type exist also in global supersymmetry. These are terms with
$m=0$ in eq.~\eqref{m3} and are present because the new FI term can be written in global supersymmetry too. In superfield notations
it reads
\beq
\mbox{New FI term}= \Xi \int d^4 \theta (D_\alpha W^\alpha + \bar{D}^{\dot\alpha} \bar{W}_{\dot\alpha} ) 
{ W^2 \bar{W}^2 \over  D^2 W^2 \bar{D}^2 \bar{W}^2} .
\eeq{m4}
The second set of terms decouples in the limit $M_{Pl}\rightarrow \infty$, $D=\Xi=constant$ so it is weighted by strictly negative
powers of $M_{Pl}$. Equation~\eqref{m3} shows that for $\Xi \lesssim M_{Pl}$ the strongest limit on the UV cutoff of the effective
theory comes from the $m=0$ terms and is
\beq
\Lambda_{cut}^2 \lesssim D=\Xi,
\eeq{m5}
where $\Lambda_{cut}$ is the cutoff scale of the theory. If other mass parameters $m \lesssim \sqrt{\Xi}$ exist in the theory, they can only appear in the numerator of~\eqref{m3} and so they only make 
the constraint on the cutoff weaker.
The scalar potential of the KKLT model has a minimum at $V=-3H^2/8\pi G$ and it asymptotes to $V=0$ for large values of the would
 be inflaton. To get a realistic inflationary potential, the D-term contribution must lift the minimum to $-3H^2 /8\pi G + \xi^2/2 \approx 0$
hence $D=\Xi \sim H M_{Pl}$. Since the D-term breaks supersymmetry we also get $D\sim M_S^2$, hence the bounds already given in 
equation~\eqref{m2}~\footnote{We
thank A. Guillen and F. Rondeau for a useful email exchange on this point.} 
\beq
\Lambda_{cut} \lesssim \sqrt{\Xi}\sim M_S, \qquad \Lambda_{cut} \lesssim \sqrt{H M_{Pl}}.
\eeq{m6}

\section{Starobinsky-type inflation}\label{infla}

In this section, we use the new FI term to derive an inflationary potential and explore its hidden-sector dynamics. Let us begin with the general potential $V=V_h+V_{soft}$, which is given by
\begin{eqnarray}
 V &=& \Big( \frac{1}{2}g^2\xi^2 + g^2\xi U^h  + \frac{1}{2}g^2{U^h}^2 \Big)  - \frac{1}{X^2}\bigg(-2aA^2e^{-a(X+\Phi/3)}+2acA^2 e^{-a(X+\Phi/3)/2} \cos(a\textrm{Im}T)\bigg)
\nonumber\\&&+ \frac{1}{3X^{2}}\Big(X+\frac{1}{3}\Phi_i\Phi^{i\bar{j}}\Phi_{\bar{j}}\Big) a^2A^2 e^{-a(X+\Phi/3)} + V_{soft}.
\end{eqnarray}
Now we assume that the hidden-sector part of the real function, say $U^h$, is defined by
\begin{eqnarray}
 U^h \equiv C_iz^i\bar{z}^i,
\end{eqnarray}
where the $z^i$'s are all matter scalars appearing in the supergravity model --except the Higgs sector fields-- and the $C_i$'s are 
coupling constants. It is easy to see that the minima of the total scalar potential $V=V_h + V_{soft}$ with respect to the matter scalars $z^i$'s without Higgs ones are placed at $z^i=0$. 

To explore the inflationary trajectory in the direction of inflaton field $\phi$ (or $X \equiv e^{\sqrt{2/3}\phi}$), we focus on the path along the minima at $z^i=0$ where again $i \neq \textrm{Higgs}$; $\textrm{Im}T=0$, and $H_u^+=H_d^-=0, H_u^0=v_u/\sqrt{2},H_d^0=v_d/\sqrt{2}$ where $v_u,v_d$ are non-zero constants. Then, along the path, the total scalar potential can be written as
\begin{eqnarray}
  V|_{minima} &=& \frac{1}{2}g^2\xi^2  - \frac{1}{X^2}\bigg(-2aA^2e^{-a(X+v^2/6)}+2acA^2 e^{-a(X+v^2/6)/2} \bigg)
  \nonumber\\
  &&+ \frac{1}{3X^{2}}\Big(X+\frac{v^2}{6}\Big) a^2A^2 e^{-a(X+v^2/6)}+ V_{soft}|_{minima},
\end{eqnarray}
where we defined $v^2 \equiv v_u^2 + v_d^2$. 
We can further simplify the form of this potential using the fact that $v = 246 \textrm{GeV} \sim 10^{-16} M_{pl} \ll X \sim \mathcal{O}(M_{pl})$ all the time during and after inflation. That is, we can take the limits $X \gg v^2/6$ and $V^h \gg V_{soft}$ during and after inflation, which produce
\begin{eqnarray}
   V|_{minima} \approx \frac{1}{2}g^2\xi^2  - \frac{1}{X^2}\bigg(-2aA^2e^{-aX}+2acA^2 e^{-aX/2} \bigg)+ \frac{1}{3X} a^2A^2 e^{-aX}.
\end{eqnarray}
The vacuum with respect to the direction $X$ is at $X=x$ such that $c=(1+ax/3)e^{-ax/2}$ (see Ref. \cite{jp2} for the derivation of $c$). In fact, the scale of $g^2\xi^2$ must be of order of the inflation energy since we want to describe inflation using that potential. That is, we must require 
\begin{eqnarray}
 \frac{1}{2}g^2\xi^2 \overset{!}{=} M_I^4 \equiv H^2M_{pl}^2,
\end{eqnarray}
where $M_I$ and $H$ denote the inflation and Hubble scale respectively. Using $X=e^{\sqrt{2/3}\phi}$, we rewrite the potential as
\begin{eqnarray}
   V|_{minima} \approx M_I^4  - e^{-2\sqrt{2/3}\phi}\bigg(-2aA^2e^{-ae^{\sqrt{2/3}\phi}}+2acA^2 e^{-ae^{\sqrt{2/3}\phi}/2} \bigg) + \frac{1}{3}a^2A^2e^{-\sqrt{2/3}\phi}  e^{-ae^{\sqrt{2/3}\phi}}.
\end{eqnarray}
It is worth noticing that this result exactly coincides with that of Ref. \cite{jp2} so that an exponentially flat direction similar to that 
of the Starobinsky potential is present in our model too. 

Our model has enough parameters to fix the post-inflation cosmological constant to the observed value $\Lambda \sim 10^{-120}M_{pl}$. Now considering the value of the soft potential at the {\it vacuum} when $X=x$, $i \neq \textrm{Higgs}$; $\textrm{Im}T=0$, and $H_u^+=H_d^-=0, H_u^0=v_u/\sqrt{2},H_d^0=v_d/\sqrt{2}$, we can determine what the constant $g^2\xi^2/2$ must be. At the vacuum, if we define the vacuum with respect to $X$ (or $\phi$) to be at $x=1$ (or $\phi=0$), the potential is given by
\begin{eqnarray}
 V|_{vacua} = \frac{1}{2}g^2\xi^2-\frac{a^2A^2e^{-a}}{3}+ \Lambda_{soft} \equiv \Lambda,
\end{eqnarray}
where we define $\Lambda_{soft} = \left<V_{soft}\right>$ and impose that the VEV of the potential is equal to the cosmological constant $\Lambda$. Hence, we determine $g^2\xi^2/2$ as
\begin{eqnarray}
 \frac{1}{2}g^2\xi^2 = \frac{a^2A^2e^{-a}}{3}+\Lambda- \Lambda_{soft}.
\end{eqnarray}

Now let us investigate supersymmetry (SUSY) breaking in our model. The SUSY breaking scale, say $M_S$, can be found by computing  the positive contributions to both D and F terms
\begin{eqnarray}
 V_+|_{vacuum} = (V + 3e^G)|_{vacua} = \Lambda + \frac{a^2A^2e^{-a}}{3} \equiv M_S^4,
\end{eqnarray}
which gives
\begin{eqnarray}
 \frac{a^2A^2e^{-a}}{3} = M_S^4 - \Lambda \implies \frac{1}{2}g^2\xi^2 = M_S^4 - \Lambda_{soft} = M_I^4 \implies M_S^4 = M_I^4 + \Lambda_{soft}.
\end{eqnarray}
This means that we have to require a high-scale supersymmetry breaking \cite{HighSUSY} because the SUSY breaking mass $M_S$ is at high scale as given by
\begin{eqnarray}
 M_S = (H^2M_{pl}^2+\Lambda_{soft})^{1/4} \sim \mathcal{O}(\sqrt{HM_{pl}}) = 10^{-2.5}M_{pl},
\end{eqnarray}
where we note that $H^2M_{pl}^2 \gg \Lambda_{soft}$.

\section{Masses and Mass Splittings}\label{mass}

In this section, we embed the minimal supersymmetric standard model (MSSM) into the observable sector of our supergravity 
theory, whose hidden sector we showed to describes both inflation and the post-inflationary vacuum.

\subsection{Supersymmetric Higgs potential modified by new FI terms}

Here we focus on finding a supersymmetric Higgs potential compatible with MSSM phenomenology in our supergravity model of inflation. To generate the observed Higgs and matter masses, we assume that the generic function $U=U^h+U^o$ is defined by
\begin{eqnarray}
 U^h &=& C_i|z^i|^2\quad\textrm{for non-Higgs matters, labeled by}~i, \\
 U^o &=& b[(|H_u^+|^2+|H_u^0|^2) - (|H_d^0|^2+|H_d^-|^2)] \quad\textrm{for the Higgs sector},
\end{eqnarray} 
where $C_i,b$ are free parameters. Notice that these are gauge invariant under the SM gauge groups. We can then identify the supersymmetric Higgs potential with the soft potential, which is given by
\begin{eqnarray}
 V_{soft} &=& \xi g^2U^o + \frac{g^2}{2}{U^o}^2 + V_{U(1)_Y}+V_{SU(2)_L} + V_{SU(3)_c} \nonumber\\
 &&+ \frac{2}{3}\frac{aAe^{-a(X+\Phi/3)}/2}{X^2}(W^o+\bar{W}^o) + \frac{1}{9}\frac{a^2A^2e^{-a(X+\Phi/3)}}{X^2}\Phi + \frac{W_i^o\delta^{i\bar{j}}\bar{W}^o_{\bar{j}}}{X^2}.
\eea{h-pot}

Since the Higgs multiplets transform under $U(1)_Y$ and $SU(2)_L$, the first line of the potential in eq.~\eqref{h-pot} is 
\begin{eqnarray}
 g^2\xi U^o &=&  g^2\xi b(|H_u^+|^2+|H_u^0|^2 - |H_d^0|^2-|H_d^-|^2)  ,\\
 \frac{g^2}{2}{U^o}^2 &=& \frac{g^2b^2}{2}(|H_u^+|^2+|H_u^0|^2 - |H_d^0|^2-|H_d^-|^2)^2  ,\\
 V_{U(1)_Y} &\supset& \frac{g_1^2}{8X^2} (|H^+_u|^2+|H^0_u|^2-|H_d^0|^2-|H_d^-|^2)^2,\\
 V_{SU(2)_L} &\supset& \frac{g_2^2}{2X^2}|\bar{H}^0_uH^+_u+\bar{H}_d^-H_d^0|^2 + \frac{g_2^2}{8X^2} (|H^+_u|^2+|H^0_u|^2-|H_d^0|^2-|H_d^-|^2)^2.
\end{eqnarray}
In addition, we can find the other part of the Higgs potential from the F-term contributions to the soft potential $ V_{soft}|_F $, which provides
\begin{eqnarray}
V_{soft}|_F &\supset& \frac{2}{3}\frac{aAe^{-a(X+\Phi/3)}/2}{X^2}(W^o+\bar{W}^o) + \frac{1}{9}\frac{a^2A^2e^{-a(X+\Phi/3)}}{X^2}\Phi + \frac{W_i^o\delta^{i\bar{j}}\bar{W}^o_{\bar{j}}}{X^2}\nonumber\\
&=& \frac{2}{3}\frac{aAe^{-a(X+\Phi/3)}/2}{X^2} \Big( \mu(H_u^+H^-_d -H^0_u H^0_d )+ h.c.\Big)\nonumber\\
&&+\bigg(\frac{1}{9}\frac{a^2A^2e^{-a(X+\Phi/3)}}{X^2}+\frac{|\mu|^2}{X^2}\bigg)(|H^+_u|^2+|H^0_u|^2+|H_d^0|^2+|H_d^-|^2)
\end{eqnarray}

Therefore, the final form of the Higgs potential at the non-Higgs matter minima $z^i=0$ (where $i\neq$ Higgs) is given by
\begin{eqnarray}
 V_H &=&\frac{g_2^2}{2X^2}|\bar{H}^0_uH^+_u+\bar{H}_d^-H_d^0|^2 + \Big(\frac{g_1^2+g_2^2}{8X^2} + \frac{g^2b^2}{2}\Big) (|H^+_u|^2+|H^0_u|^2-|H_d^0|^2-|H_d^-|^2)^2\nonumber\\
 &&+\frac{4}{3}\frac{aAe^{-a(X+(|H^+_u|^2+|H^0_u|^2+|H_d^0|^2+|H_d^-|^2)/3)}/2}{X^2} \textrm{Re}\Big(\mu(H_u^+H^-_d -H^0_u H^0_d)\Big)\nonumber\\
&&+\bigg(\frac{1}{9}\frac{a^2A^2e^{-a(X+(|H^+_u|^2+|H^0_u|^2+|H_d^0|^2+|H_d^-|^2)/3)}}{X^2}+\frac{|\mu|^2}{X^2} + g^2\xi b \bigg)(|H^+_u|^2+|H^0_u|^2)\nonumber\\
&& +\bigg(\frac{1}{9}\frac{a^2A^2e^{-a(X+(|H^+_u|^2+|H^0_u|^2+|H_d^0|^2+|H_d^-|^2)/3)}}{X^2}+\frac{|\mu|^2}{X^2}-g^2\xi b \bigg)(|H_d^0|^2+|H_d^-|^2) .
\end{eqnarray}
Since $X \gg H^{\pm,0}_{u,d}$ the potential can be approximated into
\begin{eqnarray}
  V_H &=&\frac{g_2^2}{2X^2}|\bar{H}^0_uH^+_u+\bar{H}_d^-H_d^0|^2 + \Big(\frac{g_1^2+g_2^2}{8X^2} + \frac{g^2b^2}{2}\Big) (|H^+_u|^2+|H^0_u|^2-|H_d^0|^2-|H_d^-|^2)^2\nonumber\\
 &&+\frac{4}{3}\frac{aAe^{-aX/2}}{X^2} \textrm{Re}\Big(\mu(H_u^+H^-_d -H^0_u H^0_d)\Big)\nonumber\\
&&+\bigg(\frac{1}{9}\frac{a^2A^2e^{-aX}}{X^2}+\frac{|\mu|^2}{X^2} +g^2\xi b \bigg)(|H^+_u|^2+|H^0_u|^2)\nonumber\\
&& +\bigg(\frac{1}{9}\frac{a^2A^2e^{-aX}}{X^2}+\frac{|\mu|^2}{X^2}-g^2\xi b \bigg)(|H_d^0|^2+|H_d^-|^2)
\end{eqnarray}
We then find the minima at $H_u^+=H_d^-=0$, which gives
\begin{eqnarray}
  V_H &=& \Big(\frac{g_1^2+g_2^2}{8X^2} + \frac{g^2b^2}{2}\Big) (|H^0_u|^2-|H_d^0|^2)^2-\frac{4}{3}\frac{aAe^{-aX/2}}{X^2}\textrm{Re}(\mu  H^0_u H^0_d)\nonumber\\
&&+\bigg(\frac{1}{9}\frac{a^2A^2e^{-aX}}{X^2}+\frac{|\mu|^2}{X^2} +g^2\xi b \bigg)|H^0_u|^2 +\bigg(\frac{1}{9}\frac{a^2A^2e^{-aX}}{X^2}+\frac{|\mu|^2}{X^2}-g^2\xi b \bigg)|H_d^0|^2.
\end{eqnarray}
In terms of the approximated potential, the vacuum solutions are those of the MSSM. That is, 
\begin{eqnarray}
 \left<H^{0}_{u}\right> = \frac{v_{u}}{\sqrt{2}}(= v_2),\quad \left<H^{0}_{d}\right> = \frac{v_{d}}{\sqrt{2}}(= v_1), \quad \left<H^{+}_{u}\right>= \left<H^{-}_{d}\right> = 0 \implies H^0_i \approx \left<H^{0}_{i}\right> + \varphi_i,
\end{eqnarray}
where $\varphi_i$ are fluctuations of the Higgs fields $H_i^0$ around the vacuum ($i=u,d$). We take here the same definitions used in the MSSM
\begin{eqnarray}
 v^2 \equiv v_u^2 + v_d^2 = (246~\textrm{GeV})^2, \quad \tan\beta = v_2/v_1= v_u/v_d,
\end{eqnarray}
where $\beta$ is a free parameter such that $0\leq \beta \leq \pi/2$. Hence, we can merely recall the MSSM results when we compute scalar masses. 

Recalling some results of Sec. 28.5 in Ref. \cite{WeinbergSUSY}, we can identify the following correspondences 
\begin{eqnarray}
 && \frac{g^2+g'^2}{8} \rightarrow \frac{g_1^2+g_2^2}{8X^2} + \frac{g^2b^2}{2}, \quad  m_1^2 \rightarrow \frac{1}{9}\frac{a^2A^2e^{-aX}}{X^2} -g^2\xi b, \quad  m_2^2 \rightarrow \frac{1}{9}\frac{a^2A^2e^{-aX}}{X^2} +g^2\xi b,\\
 && |\mu|^2  \rightarrow \frac{|\mu|^2}{X^2}, \quad  B\mu \rightarrow \frac{4}{3}\frac{aAe^{-aX/2}}{X^2}\mu , \\
 && m_Z^2 = \frac{1}{2}(g^2+g'^2)(v_1^2+v_2^2) \rightarrow m_Z'^2 =\left(\frac{g_1^2+g_2^2}{2X^2} + 2g^2b^2\right)(v_1^2+v_2^2)= \left(X^{-2}+\frac{4g^2b^2}{g_1^2+g_2^2}\right)m_Z^2 ,\nonumber\\{}\\
 && m_A^2 = 2|\mu|^2 + m_1^2 + m_2^2 \rightarrow m_A'^2 =  \frac{2}{9}\frac{a^2A^2e^{-aX}}{X^2}+\frac{2|\mu|^2}{X^2},
\end{eqnarray}
and the vacuum solutions produces the following relations
\begin{eqnarray}
 B\mu = {m'}_A^2 \sin 2\beta, \quad m_1^2 - m_2^2 = - ({m'}_A^2+{m'}_Z^2)\cos 2\beta = -2g^2\xi b, \quad \tan \beta = v_2/v_1. 
\end{eqnarray}

\subsection{Supermassive scalars}

The scalar masses are determined as follows. The Standard Model matter masses are found to be 
\begin{eqnarray}
 m_z^2|_{vac}  &=& V_{z\bar{z}}|_{z=0} =  \Big(g^2\xi U_{z\bar{z}} + g^2(U_zU_{\bar{z}}+UU_{z\bar{z}}) + (V_F + V_D')_{z\bar{z}}\Big)\Big|_{z=0}
 \nonumber \\
 &=& (g^2\xi U_{z\bar{z}}+ (V_F + V_D')_{z\bar{z}})|_{z=0} \gg H^2 \nonumber\\
 &\implies& g^2\xi U_{z\bar{z}}|_{z=0} \gg H^2
\end{eqnarray}
along the vacua when $z=0,a\textrm{Im}T=0$. Now we may suppose that  the form of the general function $U$ is 
\begin{eqnarray}
 U \supset  C_{i} z^i\bar{z}^i,
\end{eqnarray}
where $z^i$'s are the matter fields without the Higgs fields. This leads to 
\begin{eqnarray}
  g^2\xi U_{z\bar{z}} = g^2\xi C_i \gg H^2 \implies C_i \gg \frac{H^2}{g^2\xi}>0.
\end{eqnarray}
Notice that $U$ is positive definite, so that $D = \xi + U >0$ is nowhere vanishing. Here the point is that the matter scalars can be as mush heavy as we want during and after inflation, enabling us to integrate out them easily. 
 
Next, let us identify the Higgs masses. The eigenvalues of the square mass matrix of the Higgs fields are
\begin{eqnarray}
 m_H^2 &=& \frac{1}{2}({m'}_A^2+{m'}_Z^2 + \sqrt{({m'}_A^2+{m'}_Z^2)^2 - 4{m'}_A^2{m'}_Z^2 \cos^22\beta}), \label{higgs-eig1}\\
  m_h^2 &=& \frac{1}{2}({m'}_A^2+{m'}_Z^2 - \sqrt{({m'}_A^2+{m'}_Z^2)^2 - 4{m'}_A^2{m'}_Z^2 \cos^22\beta}) \approx \frac{{m'}_A^2{m'}_Z^2\cos^2 2\beta}{{m'}_A^2+{m'}_Z^2} \nonumber\\
  &\approx& \frac{(m_1^2-m_2^2)^2}{{m'}_A^2{m'}_Z^2} \quad \textrm{if}\quad {m'}_A \gg {m'}_Z \implies {m'}_A^2 \approx  \frac{(m_1^2-m_2^2)^2}{m_h^2{m'}_Z^2} 
\eea{higgs-eig2}
where $\mu, B,m_1^2,m_2^2$ are the MSSM soft parameters. We note that now the MSSM soft parameters are functions of the inflaton field $\phi$ via $X \equiv e^{\sqrt{2/3}\phi}$, whose vacuum is at $\phi=0$ (or $X=1$). First, let us check the Higgs masses after inflation at $X=1$. Eqs~(\ref{higgs-eig1},\ref{higgs-eig2})  imply that
\begin{eqnarray}
 {m'}_A^2 = \frac{2a^2A^2e^{-a}}{9} + 2|\mu|^2 \approx \frac{(m_1^2-m_2^2)^2}{m_h^2{m'}_Z^2}, 
\end{eqnarray}
so that we can determine the parameter $\mu$ as
\begin{eqnarray}
 |\mu|^2  \approx \frac{(m_1^2-m_2^2)^2}{2m_h^2{m'}_Z^2} - \frac{1}{3}(M_I^4-\Lambda) 
\end{eqnarray}
since $3(M_I^4-\Lambda) = a^2A^2e^{-a} = 3m_{3/2}^2$. Since we have 
\begin{eqnarray}
 && m_1^2 - m_2^2 = -2g^2\xi b \approx -2g^2 b \frac{\sqrt{2}M_I^2}{g} \sim 2\sqrt{2}bgM_I^2,
\end{eqnarray}
we obtain
\begin{eqnarray}
 && |\mu|^2  \sim  \bigg( \frac{4 b^2g^2}{m_h^2m_Z^2 \left(1+\frac{4g^2b^2}{g_1^2+g_2^2}\right)} -\frac{1}{3} \bigg)  M_I^4 >0,
\end{eqnarray}
where $m_Z^2 = \frac{g_1^2+g_2^2}{4} v^2$ is the Z boson mass.

We remark that it is necessary to consider the new FI term in this model since it helps us to obtain different values for $m_1^2$ and $m_2^2$. This property ensures that the light Higgs scalar mass $m_h$ is non-vanishing. Furthermore, we observe that we can integrate out the degrees of freedom of the heavy Higgs scalar, with mass $m_H^2$, because this mass is of order of the Hubble scale, while the light Higgs mass can be fixed to be that of the observed Higgs using the cancellation between the first and second terms in the mass formula. Next, let us inspect the Higgs masses during inflation for $X \gg 1$. In this phase, we have
\begin{eqnarray}
 {m'}_A^2 \rightarrow 0, \quad {m'}_Z^2 \rightarrow \frac{4g^2b^2}{g_1^2+g_2^2}m_Z^2 \implies m_H^2 \rightarrow  \frac{4g^2b^2}{g_1^2+g_2^2}m_Z^2, \quad m_h^2 \rightarrow 0.
\end{eqnarray}
We thus need to impose 
\begin{eqnarray}
 \frac{4g^2b^2}{g_1^2+g_2^2}m_Z^2 =  \frac{g^2b^2}{v^2} \gg H^2 \implies g \gg \frac{vH}{b}.\label{coupling_constraint}
\end{eqnarray}
Since we have $g \sim M_S^{-2}$ and $M_S^2 \sim H \sim 10^{-5}$, Eq.~\eqref{coupling_constraint} reduces to
\begin{eqnarray}
b \gg  v H M_S^2 \sim 10^{-26},
\end{eqnarray}
We observe that the parameter $b$ corresponds to the scale of the low energy observable sector if the parameter $b$ is within 
the range $\xi \sim M_S^4 = M_I^4 = H^2 \sim 10^{-10} \gg b \gg v H M_S^2 \sim 10^{-26}$. In the limit, the $\mu$ term becomes 
\begin{eqnarray}
  |\mu|^2  \sim  \bigg( \frac{4v^2}{m_h^2} -\frac{1}{3} \bigg)  M_I^4  >0 \implies \mu \sim \mathcal{O}(H).
\end{eqnarray}
Hence, we need to obey the constraint 
\begin{eqnarray}
 v > \frac{m_h}{2\sqrt{3}},
\end{eqnarray}
which can be satisfied since we already have $v>m_h$ with the observed values, $v = 246~\textrm{GeV}$ and $m_h = 125~\textrm{GeV}$. The impact of quantum loop corrections on the Higgs mass is an important topic that we leave  for future investigations.

We now summarize the spectra of the scalar masses. We find that only the light Higgs scalar mass $m_h$ varies from almost zero during inflation to the observed Higgs mass $m_h \sim 125~\textrm{GeV}$ at the true vacuum after inflation. On the other hand, the other scalar masses in this model can be much heavier than the Hubble scale during and after inflation, so  they do not contribute to the dynamics of slow-roll inflation. 

As for the light Higgs mass during inflation this seems to be a problem for single-field slow roll inflation at first glance. However, according to Ref.~\cite{LightScalar}, it is possible to have a robust slow-roll inflation even when extra light scalars are present if some reheating scenario conditions are satisfied. We find that our model may be allowed to satisfy either ``Case-5'' or ``Case-8'' reheating scenarios of~\cite{LightScalar}, which are strongly favored according to~\cite{LightScalar}. The corresponding conditions are as follows:
\begin{eqnarray}
&& \textrm{Case-5}:~ \Gamma_{h} < \Gamma_{\phi} < m_h < H, \qquad \left(\frac{\Gamma_h}{\Gamma_{\phi}}\right)^{1/4}\ll \frac{\left<h\right>}{M_{pl}}\sim \frac{v}{M_{pl}} \ll 1,\\
&& \textrm{Case-8}:~ \Gamma_{h} < m_h < \Gamma_{\phi} < H, \qquad \left(\frac{\Gamma_h}{m_h}\right)^{1/4}\ll \frac{\left<h\right>}{M_{pl}}\sim \frac{v}{M_{pl}} \ll 1,
\end{eqnarray}
where $\Gamma_{\phi},\Gamma_h$ are the decay rates of the inflaton $\phi$ and  the light Higgs $h$ during the reheating phase, and $v$ is the VEV of the Higgs after inflation. Note that the decay rate of Higgs has to be the smallest.

We also note that unlike our previous model in Ref. \cite{jp2}, we can specify the reheating scenario conditions using the observed values for $m_h$ and $v$\footnote{i.e. $v=246~\textrm{GeV} \sim 10^{-16}M_{pl}$ and $m_h = 125~\textrm{GeV} \sim 10^{-16} M_{pl}$ while $H \sim 10^{-5}M_{pl}$} and make use of them to constrain the decay rates. We leave to future work a detailed
 study of the decay rates and and reheating scenarios.

\subsection{Ultralight SM fermions and heavy sfermions}

In this section, we compute fermionic masses in our supergravity model. First, we recall the superpotential in our model 
\begin{eqnarray}
 W(T) \equiv W^h(T)+W^o(Z^i), 
\end{eqnarray}
where
\begin{eqnarray}
 W^h(T) &\equiv& W_0 + Ae^{-aT}, \\
 W^o(Z^i) &\equiv& W_{MSSM} =  -Y_u \hat{U}_R \hat{H}_u \cdot \hat{Q} + Y_d \hat{D}_R \hat{H}_d \cdot \hat{Q} + Y_e \hat{E}_R \hat{H}_u \cdot \hat{L}  + \mu \hat{H}_u \cdot \hat{H}_d \nonumber\\
 &=&  -Y_u \tilde{u}_R (H^+_u \tilde{d}_L -H^0_u\tilde{u}_L) + Y_d \tilde{d}_R (H^0_d \tilde{d}_L -H^-_d\tilde{u}_L)\nonumber\\
 && + Y_e \tilde{e}_R (H^+_u \tilde{\nu}_L -H^0_u\tilde{e}^-_L)  + \mu (H^+_uH^-_d-H^0_uH^0_d).
\end{eqnarray}
The most general fermion masses $m^{(g)}$ are given by all the contributions from the standard supergravity, new FI terms, and the super-Higgs effects to the fermion mass, which are written in~\eqref{General_Fermion_Masses}. Here, we point out that if the gauge kinetic function is purely a constant, then the gaugino masses almost vanish at the vacuum. In particular, when a gauged R-symmetry is imposed, gauginos can get massive enough thanks to the $U_R(1)$ anomaly cancellation between one-loop quantum correction to the Lagrangian   and the shift of Green-Schwarz term by the presence of a linear term in some charged moduli in the gauge kinetic function. However, in our model, we consider a model with no gauging of the R-symmetry. Thus, we can just add a linear term in the gauge kinetic function as follows:
\begin{eqnarray}
 f_{AB}(T) = \delta_{AB}\Big( \frac{1}{\sqrt{g_Ag_B}} + \sqrt{\beta_A\beta_B} T\Big), 
\end{eqnarray}
where $T$ is the modulus field and $\delta_{AB}$ is the Kronecker delta. We also assume that the coefficient $\beta_A$ can be sufficiently small so that
\begin{eqnarray}
 g^{-2}_A \gg \beta_A T \implies g^{-2}_A \gg \beta_A  ~~\textrm{at the vacuum where } T \sim 1,
\end{eqnarray}
so that the gauge kinetic Lagrangian  s is still approximately canonically normalized. We note that in any case we can make the scale of $\beta g^2$  very small
\begin{eqnarray}
g \equiv 10^{-n}, \beta \equiv 10^m \implies \beta g^2=10^{m-2n} \ll 1 \implies m < 2n.
\end{eqnarray}
This will be used for estimating the gaugino masses. For example, when the gauge coupling is sufficiently small, i.e. $g=10^{-n} \ll 1$, we may consider $\beta \sim \mathcal{O}(10^m)$ where $0<m<2n$. This will contribute to the fermion masses as a large number in our model. The smaller $g$ gets, the larger $\beta$ can become.

Then, the correspondimg fermion mass expressions reduce to the following
\begin{eqnarray}
 m_{3/2} &=& We^{K/2},\\
  m_{IJ}^{(g)} &=&  e^{K/2}(W_{IJ} + K_{IJ}W+K_JW_I+K_IW_J + K_IK_JW)
\nonumber\\
&& -e^{K/2} G^{K\bar{L}}\partial_I G_{J\bar{L}}(W_K + K_KW) -\frac{2}{3}  (W_I+K_IW)(W_J+K_JW),\nonumber\\
m_{IA}^{(g)} &=& 
 i\sqrt{2}[\partial_I\mathcal{P}_A-\frac{1}{4}\delta_{A}^C\sqrt{\beta_A\beta_C}\delta_{IT} \Big( \frac{1}{\sqrt{g_Ag_C}} + \sqrt{\beta_A\beta_C} \textrm{Re}T\Big)^{-1}\mathcal{P}_C]-i\frac{2}{3\sqrt{2}W}(W_I+K_IW) \mathcal{P}_A
\nonumber\\{}\\
m_{AB}^{(g)} &=&  -\frac{1}{2}e^{K/2}\delta_{AB}\sqrt{\beta_A\beta_B} G^{T\bar{J}}(\bar{W}_{\bar{J}}+K_{\bar{J}}\bar{W}) + \frac{1}{3e^{K/2}W} \mathcal{P}_A\mathcal{P}_B\\
m_{I\lambda}^{(g)} &=&  -\frac{i}{\sqrt{2}}\frac{\mathcal{U}_I}{\mathcal{U}}-\frac{i\sqrt{2}}{3W}(W_I+K_IW) \mathcal{U} = m_{\lambda I}^{(g)},\\
m_{\lambda\lambda}^{(g)} &=&  -e^{K/2} \left( \bar{W} + 4G^{I\bar{J}}\left(\frac{\mathcal{U}_I}{\mathcal{U}}+\frac{K_I}{3}\right)(\bar{W}_{\bar{J}}+K_{\bar{J}}\bar{W}) \right) +\frac{\mathcal{U}^2}{3e^{K/2}W},
\end{eqnarray}
where $\lambda^A$ is the gaugino corresponding to the gauge multiplet $V_A$ ($A=SU(3)_c,SU(2)_L,U(1)_Y$), and $\lambda$ is the superpartner of the new FI term vector multiplet $V$. Remember that $\mathcal{U}$ is nowhere vanishing by definition; that is, $\mathcal{U} = \xi + U>0$ with $U\geq 0$ and $\xi > 0$. The detailed derivation of the masses is present in the appendix. We note that the gravitino in this model has a mass $O(H)$, i.e. in the super-EeV-range. Refs. \cite{EeVGravitino,HeavyGravitino,GDM} show that
a gravitino in the EeV mass range can be a heavy dark matter candidate. 

We have checked that only the neutral components $H_u^0,H_d^0$ of the Higgs fields have non-vanishing vacuum expectation values (VEV), while all other matter scalars have vanishing VEVs. In addition, we have assumed that $\mathcal{U}|_{vac}=(\xi+U)|_{vac}>0$.  Then, we have the following vacuum expectation values at the minimum ($\left<H_a^0\right> = v_a/\sqrt{2}$ and $\left<z^{i'}\right>=0$): 
\begin{eqnarray}
 && \Phi_{a}|_{vac} = \frac{v^2_a}{2}, \quad \Phi_{i'}|_{vac}  = 0, \quad \mathcal{U}|_{vac} \approx \xi \sim M_S^4 \sim H^2, \quad X|_{vac} = 1, \quad  W^o_{i'} |_{vac} =0, \quad W^o_{i'b}|_{vac}=0 ,\nonumber\\
 && m_{3/2} = e^{G/2}|_{vac} = \sqrt{|W|^2}|_{vac} \sim H,\quad \mu \sim \mathcal{O}(H),
\end{eqnarray}
where $H_u^0,H_d^0$ are labeled by the index $a$, and the $z^i$'s, including $H_u^+,H^-_d$, are labeled by the index $i'$ (in which $i\neq a$).

The moment maps with respect to the gauge groups of SM are given by
\begin{eqnarray}
 \mathcal{P}_{U(1)_Y} &=& \frac{g_1}{X} \bigg[ \sum_{i=gen} \Big( \frac{1}{6}\tilde{Q}^{\dag}_i\tilde{Q}_i-\frac{1}{2}\tilde{L}^{\dag}_i\tilde{L}_i -\frac{2}{3} \tilde{u}^{\dag}_{R_i}\tilde{u}_{R_i} + \frac{1}{3} \tilde{d}^{\dag}_{R_i}\tilde{d}_{R_i} + \tilde{l}^{\dag}_{R_i}\tilde{l}_{R_i}
 \Big) + \frac{1}{2}H^{\dag}_u H_u -\frac{1}{2}H^{\dag}_d H_d\bigg],\nonumber\\{}\\
  \mathcal{P}_{SU(2)_L} &=&  \frac{g_2}{X} \bigg[\sum_{i=gen} \Big(\tilde{Q}^{\dag}_i\frac{\vec{\sigma}}{2}\tilde{Q}_i  +\tilde{L}^{\dag}_i\frac{\vec{\sigma}}{2}\tilde{L}_i \Big) + H^{\dag}_u \frac{\vec{\sigma}}{2} H_u+H^{\dag}_d \frac{\vec{\sigma}}{2} H_d\bigg],\\
   \mathcal{P}_{SU(3)_c} &=& \frac{g_3}{X} \bigg[ \sum_{i=gen} \Big( \tilde{Q}^{\dag}_i\frac{\vec{\lambda}}{2}\tilde{Q}_i  - \tilde{u}^{\dag}_{R_i}\frac{\vec{\lambda}}{2}\tilde{u}_{R_i}-
   \tilde{d}^{\dag}_{R_i}\frac{\vec{\lambda}}{2}\tilde{d}_{R_i}\Big)\bigg],
\end{eqnarray}
where tilded fields are superpartner scalars to the SM fermions; $\vec{\sigma}$ and $\vec{\lambda}$ are Pauli and Gell-Mann matrices; $g_1,g_2,g_3$ are gauge couplings, and the index $i$ runs over the three generations of particle in the SM. Their vacuum expectation values are
\begin{eqnarray}
 \left<\mathcal{P}_{U(1)_Y}\right> = \frac{g_1}{4}(v_u^2 - v_d^2),\quad 
 \left<\mathcal{P}_{SU(2)_L}\right> = -\frac{g_2}{4}(v_u^2 - v_d^2),\quad 
  \left<\mathcal{P}_{SU(3)_c}\right> =  0, 
\end{eqnarray}
and
\begin{eqnarray}
&& \left<\partial_{H_u^0}\mathcal{P}_{U(1)_Y}\right> = \frac{g_1v_u}{2\sqrt{2}} + \frac{g_1v_u}{12\sqrt{2}}(v_u^2 - v_d^2), \quad
 \left<\partial_{H_d^0}\mathcal{P}_{U(1)_Y}\right> = -\frac{g_1v_d}{2\sqrt{2}}+ \frac{g_1v_d}{12\sqrt{2}}(v_u^2 - v_d^2),\\
&& \left<\partial_{H_u^0}\mathcal{P}_{SU(2)_L}\right>  = - \frac{g_2v_u}{2\sqrt{2}} - \frac{g_2v_u}{12\sqrt{2}}(v_u^2 - v_d^2) , \quad 
  \left<\partial_{H_d^0}\mathcal{P}_{SU(2)_L}\right>  =  \frac{g_2v_d}{2\sqrt{2}} -\frac{g_2v_d}{12\sqrt{2}}(v_u^2 - v_d^2),
\end{eqnarray}
where $\left<\partial_{I}\mathcal{P}_{A}\right>=0$ for other scalars.

Now we are ready to estimate the scales of the fermionic masses. First, we estimate the masses of matter fermions. Given the supergravity G-function $G = -3\ln[T+\bar{T}-\Phi/3] + \ln W + \ln\bar{W}$ with the superpotential $W = W^h(T) +W^o(z^i)$, the components of the fermion mass matrix are as follows: 
\begin{eqnarray}
m_{ij}^{(g)} &=& \sqrt{\frac{1}{X^3}}\Big[W_{ij}^o
+\frac{2}{3X}(W_i^o\Phi_j+\Phi_iW_j^o)+\frac{2}{3X^2}\Phi_i\Phi_jW
\nonumber\\
&&+ \frac{2\Phi_i\Phi_j}{9X^2}(\Phi-\Phi_{m}\Phi^{m\bar{l}}\Phi_{\bar{l}})\left(\frac{W_T^h}{3}-\frac{W}{X}\right) 
\Big] - \frac{2}{3} \Big(W_i^o+\frac{\Phi_iW}{X}\Big)\Big(W_j^o+\frac{\Phi_jW}{X}\Big),\\
 m_{iT}^{(g)} &=& \sqrt{\frac{1}{X^3}}\Big[
 -\frac{W_i^o}{X} + \frac{2\Phi_i}{X}\left(\frac{W_T^h}{3}-\frac{W}{X}\right) - \left(\frac{W_T^h}{3}-\frac{W}{X}\right)(\Phi-\Phi_{m}\Phi^{m\bar{l}}\Phi_{\bar{l}})\frac{2\Phi_i}{3X^2}
 \Big] \nonumber\\
 &&-\frac{2}{3}\Big(W_i^o+\frac{\Phi_i}{X}W\Big)\Big(W_T^h-\frac{3}{X}W\Big),\\
 m_{TT}^{(g)} &=& \frac{6}{X}\left(\frac{W}{X}-\frac{W_T^h}{3}\right)\Big(
 1+\frac{1}{3X}(\Phi-\Phi_{m}\Phi^{m\bar{l}}\Phi_{\bar{l}})
 \Big),
\end{eqnarray}

If $\Phi= \delta_{i\bar{j}}z^i\bar{z}^{\bar{j}}$, then $\Phi=\Phi_{m}\Phi^{m\bar{l}}\Phi_{\bar{l}}$. Thus, the components reduce to 
\begin{eqnarray}
m_{ij}^{(g)} &=& \sqrt{\frac{1}{X^3}}\Big[W_{ij}^o
+\frac{2}{3X}(W_i^o\Phi_j+\Phi_iW_j^o)+\frac{2}{3X^2}\Phi_i\Phi_jW\Big]
 - \frac{2}{3} \Big(W_i^o+\frac{\Phi_iW}{X}\Big)\Big(W_j^o+\frac{\Phi_jW}{X}\Big),\nonumber\\{}\\
 m_{iT}^{(g)} &=& \sqrt{\frac{1}{X^3}}\Big[ 
 -\frac{W_i^o}{X} + \frac{2\Phi_i}{X}\left(\frac{W_T^h}{3}-\frac{W}{X}\right) 
 \Big] -\frac{2}{3}\Big(W_i+\frac{\Phi_i}{X}W\Big)\Big(W_T^h-\frac{3}{X}W\Big),\\
 m_{TT}^{(g)} &=& \frac{6}{X}\left(\frac{W}{X}-\frac{W_T^h}{3}\right).
\end{eqnarray}
The non-trivial components at the vacuum are then given by
\begin{eqnarray}
 m_{i'j'}^{(g)} &=& W_{i'j'}^o \approx \frac{v}{\sqrt{2}}Y_{i'j'},\\
  m_{uu}^{(g)} &=&-\frac{2}{3}\mu v_uv_d+\frac{1}{6}v_u^2W
 - \frac{2}{3} \Big(-\mu \frac{v_d}{\sqrt{2}}+\frac{v_u^2W}{2}\Big)^2\approx \mu v^2 \sim \mathcal{O}(H v^2) \sim m_{dd}^{(g)},\\
  m_{ud}^{(g)} &=& \Big[W_{ud}^o
+\frac{1}{3}(W_u^ov_d+v_uW_d^o)+\frac{1}{6}v_uv_dW\Big]
 - \frac{2}{3} \Big(W_u^o+\frac{v_u^2W}{2}\Big)\Big(W_d^o+\frac{v_d^2W}{2}\Big) \nonumber\\
 &\approx& -\mu \sim -H \sim -m_{+-}^{(g)},\\
  m_{uT}^{(g)} &=& \Big[ 
 -W_u^o + v_u^2\left(\frac{W_T^h}{3}-W\right)\Big] -\frac{2}{3}\Big(W_u+\frac{v_u^2}{2}W\Big)\Big(W_T^h-3W\Big) \approx \mu \frac{v_d}{\sqrt{2}}\sim Hv,\\
   m_{dT}^{(g)} &=& \Big[ 
 -W_d^o + v_d^2\left(\frac{W_T^h}{3}-W\right) 
 \Big] -\frac{2}{3}\Big(W_d+\frac{v_d^2}{2}W\Big)\Big(W_T^h-3W\Big)\approx \mu \frac{v_u}{\sqrt{2}} \sim Hv,\\
 m_{TT}^{(g)} &=& 6W-2W_T^h \approx m_{3/2} \sim H,
\end{eqnarray}
where $i'$' denotes non-Higgs matters, and the indices $\pm$ denote $H^+_u$ and $H^-_d$ respectively. Next, let us estimate the other mass parameters. We find
\begin{eqnarray}
 m_{uB}^{(g)} \sim m_{dB}^{(g)} &\approx& g_B(iv - i \frac{(-\mu v + v H)}{H} v^2 ) \sim ig_Bv,\\
m_{AB}^{(g)} &\approx& g_Ag_B\frac{v^2}{H} - H \delta_{AB}\sqrt{\beta_A\beta_B} \sim -\mathcal{O}(\beta H), \\
m_{AT}^{(g)} &\approx& g_A\mathcal{O}(v^2)- \beta_A g_A^2 \left<\mathcal{P}^A\right>,\\
m_{A\lambda}^{(g)} &\approx& g_A\mathcal{O}(v^2),\\
m_{u\lambda}^{(g)} &\approx& -i\frac{bv_u}{\xi} -i \frac{-\mu v_d+v_uH}{H} \xi \sim  -i\frac{bv}{H} \sim m_{d\lambda}^{(g)},\\
m_{T\lambda}^{(g)} &\approx&  \xi \sim H,\\
m_{\lambda\lambda}^{(g)} &\approx& m_{3/2} + \frac{\xi^2}{m_{3/2}} \sim H + \frac{H^2}{H} \sim H ,
\end{eqnarray}
where $A,B = 1,2$ for $U(1)_Y$ and $SU(2)_L$ respectively. In terms of $m_{u\lambda},m_{d\lambda}$ and since $10^{-26} \ll b \ll 10^{-10}$, we have 
\begin{eqnarray}
 10^{-32} \ll m_{u\lambda}, m_{d\lambda} \ll 10^{-16} \ll H.
\end{eqnarray}

In summary, the fermion mass matrix in the post-inflationary vacuum is 
\begin{eqnarray}
 M_f &\equiv& 
 \begin{pmatrix}
 m_{i'j'}^{(g)} &  m_{i'u}^{(g)} &  m_{i'd}^{(g)} &  m_{i'+}^{(g)} &  m_{i'-}^{(g)} & m_{i'T}^{(g)} & m_{i'B}^{(g)}  & m_{i'\lambda}^{(g)} \\
 m_{uj'}^{(g)} &  m_{uu}^{(g)} &  m_{ud}^{(g)} &  m_{u+}^{(g)} &  m_{u-}^{(g)} &  m_{uT}^{(g)} & m_{uB}^{(g)}  & m_{u\lambda}^{(g)} \\
 m_{dj'}^{(g)} &  m_{du}^{(g)} &  m_{dd}^{(g)} &  m_{d+}^{(g)} &  m_{d-}^{(g)} &  m_{dT}^{(g)} & m_{dB}^{(g)}  & m_{d\lambda}^{(g)} \\
 m_{+j'}^{(g)} &  m_{+u}^{(g)} &  m_{+d}^{(g)} &  m_{++}^{(g)} &  m_{+-}^{(g)} &  m_{+T}^{(g)} & m_{+B}^{(g)}  & m_{+\lambda}^{(g)} \\
 m_{-j'}^{(g)} &  m_{-u}^{(g)} &  m_{-d}^{(g)} &  m_{-+}^{(g)} &  m_{--}^{(g)} &  m_{-T}^{(g)} & m_{-B}^{(g)}  & m_{-\lambda}^{(g)} \\
 m_{Tj'}^{(g)} &  m_{Tu}^{(g)} &  m_{Td}^{(g)} &  m_{T+}^{(g)} &  m_{T-}^{(g)}  &  m_{TT}^{(g)} & m_{TB}^{(g)}  & m_{T\lambda}^{(g)} \\
 m_{Aj'}^{(g)} &  m_{Au}^{(g)} &  m_{Ad}^{(g)} &  m_{A+}^{(g)} &  m_{A-}^{(g)} &  m_{AT}^{(g)} & m_{AB}^{(g)}  & m_{A\lambda}^{(g)} \\
 m_{\lambda j'}^{(g)} &  m_{\lambda u}^{(g)} &  m_{\lambda d}^{(g)} &  m_{\lambda +}^{(g)} &  m_{\lambda -}^{(g)}  &  m_{\lambda T}^{(g)} & m_{\lambda B}^{(g)}  & m_{\lambda\lambda}^{(g)} \\
 \end{pmatrix}
 \nonumber\\
 &\approx&  
 \begin{pmatrix}
\frac{v}{\sqrt{2}}Y_{ij} & 0 &  0 &  0 &  0 & 0 & 0  & 0 \\
0 &  \mathcal{O}(Hv^2) &   -H &  0 &  0 & Hv & ivg_B  & -i\frac{bv}{H}\\
0 &  -H & \mathcal{O}(Hv^2)  & 0 &  0 &  Hv & ivg_B  & -i\frac{bv}{H} \\
0 & 0 & 0 & 0 & H & 0 & 0 & 0 \\
0 & 0 & 0 & H & 0 & 0 & 0 & 0 \\
0 & Hv &  Hv &  0 &  0 & H & \mathcal{O}(v^2)g_B & H \\
 0 & ivg_A &  ivg_A &  0 &  0 & \mathcal{O}(v^2)g_A & -\mathcal{O}(\beta H) & 0 \\
0& -i\frac{bv}{H} &  -i\frac{bv}{H} &  0 &  0 & H & 0  & H \\
 \end{pmatrix} \nonumber\\{}
\end{eqnarray}
where $m_{A\lambda}^{(g)}=m_{\lambda B}^{(g)} =0$ since there are no couplings between the relevant vector multiplets. Keeping the Yukawa masses of the matter fermions and dropping all other terms much smaller than the Hubble scale, the fermion mass matrix can be approximated by
\begin{eqnarray}
 M_f \approx 
  \begin{pmatrix}
\frac{v}{\sqrt{2}}Y_{ij} & 0 &  0 & 0 &  0 & 0 & 0  & 0 \\
0 &  0 &   -H & 0 & 0 &  0 &0  & 0\\
0 &  -H & 0  &  0 & 0 &  0 &0  & 0 \\
0 & 0 & 0 & 0 & H & 0 & 0 & 0 \\
0 & 0 & 0 & H & 0 & 0 & 0 & 0 \\
0 & 0 & 0 & 0 &  0 & H & 0 & H \\
0 & 0 & 0 & 0 &  0 & 0 & -\beta H  & 0 \\
0 & 0 & 0 & 0 &  0 & H & 0  & H \\
 \end{pmatrix}
\end{eqnarray}
The masses of the SM matter fermions can be matched with the observed values by adjusting the Yukawa couplings which are free parameters. {By diagonalizing the fermion mass matrices may produce negative (positive) mass eigenvalues, but the masses can be made to be positive (negative) by absorbing the sign into the mixing matrices that get imaginary (i.e. by a chiral rotation) \cite{Fermion_mass_matrix}.} We note that the chargino, neutralino, and gaugino masses at the true vacua after inflation are of the order of Hubble scale $\mathcal{O}(10^{-5})M_{pl} \sim \mathcal{O}(10^{13})~\textrm{GeV}$, implying that they may be candidates for the so-called supermassive dark matter ``WIMPZILLA'' \cite{WIMPZILLA1,WIMPZILLA2,WIMPZILLA3,WIMPZILLA4} (or superheavy dark matter in Ref. \cite{SuperheavyDM}). 

Finally, we can summarize all the parameters in our supergravity model of inflation compatible with MSSM as follows:
\begin{itemize}
    \item Hubble Scale $H$ ($\sim \mu, g^{-1}, M_S^2,A$) for inflation,
    \item Yukawa couplings $Y_{i'j'}$ for fermion masses,
    \item Neutral Higgs VEVs $v_u,v_d$ such that $v \equiv \sqrt{v_u^2 + v_d^2}$ for Higgs mechanism, which also determines the angle between $v_u$ and $v_d$, i.e. $\tan \beta = v_u/v_d$,
    \item Gauge couplings $g_1,g_2,g_3$ for strong, weak, and hypercharge interactions in the SM,
    \item New-FI-term hidden-sector parameters $C_i$'s for producing scalar bosons heavier than Hubble,
    \item New-FI-term observable-sector parameter $b$ for generating supersymmetric Higgs potential,
    \item Hidden-sector superpotential parameter $a$ for KKLT superpotential.
    \item Gauge kinetic term parameters $\beta_A$, to make gauginos massive.
\end{itemize}

%Moreover, since our supergravity model is string-motivated, it is crucial to determine whether some possible extensions of our theory belongs to string landscape \cite{Swampland} (see Refs. \cite{Recent_Review_Swampland1,Recent_Review_Swampland2} for a recent review of swampland conjectures).  

%\section{Conclusions and future directions} 
%Gauging, inclusion of matter, decorations. Very brief.

\subsection*{Acknowledgments} 
We would like to thank Anthony Guillen and Fran\c{c}ois Rondeau for useful discussions. M.P.\ is supported in part by NSF grant PHY-1915219. M.P. would like to thank Imperial College, London, for its kind hospitality during the completion of this paper. H.J. was supported by  a James Arthur Graduate Associate Fellowship (JAGA).

\setcounter{equation}{0}
\renewcommand{\theequation}{A.\arabic{equation}} 

\appendix
\section{Derivation of fermion masses in superconformal tensor calculus}

We consider matter chiral multiplets $Z^i$, the chiral compensator $S_0$, a real multiplet $V$, and another real multiplet 
$(V)_D$, whose lowest component is the auxiliary D term of the real multiplet $V$. Their superconformal multiplets are given as follows:
\begin{eqnarray}
&& V = \{0,0,0,0,A_{\mu},\lambda,D\} ~\textrm{in the Wess-Zumino gauge,~i.e.}~v=\zeta=\mathcal{H}=0, \\
&& Z^i = (z^i,-i\sqrt{2}P_L\chi^i,-2F^i,0,+i\mathcal{D}_{\mu}z^i,0,0) = \{ z^i, P_L\chi^i,F^i\},\\
&& \bar{Z}^{\bar{i}} = (\bar{z}^{\bar{i}},+i\sqrt{2}P_R\chi^{\bar{i}},0,-2\bar{F}^{\bar{i}},-i\mathcal{D}_{\mu}\bar{z}^{\bar{i}},0,0) = \{ \bar{z}^{\bar{i}}, P_R\chi^{\bar{i}},\bar{F}^{\bar{i}}\},\\
&& S_0 = (s_0,-i\sqrt{2}P_L\chi^0,-2F_0,0,+i\mathcal{D}_{\mu}s_0,0,0) = \{ s_0, P_L\chi^0,F_0\},\\
&& \bar{S}_0 = (\bar{s}_0,+i\sqrt{2}P_R\chi^0,0,-2\bar{F}_0,-i\mathcal{D}_{\mu}\bar{s}_0,0,0) = \{\bar{s}_0, P_R\chi^0,\bar{F}_0\},\\
&& \bar{\lambda}P_L\lambda = (\bar{\lambda}P_L\lambda,-i\sqrt{2}P_L\Lambda,2D_-^2,0,+i\mathcal{D}_{\mu}(\bar{\lambda}P_L\lambda),0,0) = \{\bar{\lambda}P_L\lambda, P_L\Lambda,-D_-^2\},\\
&& \bar{\lambda}P_R\lambda = (\bar{\lambda}P_R\lambda,+i\sqrt{2}P_R\Lambda,0,2D_+^2,-i\mathcal{D}_{\mu}(\bar{\lambda}P_R\lambda),0,0) = \{\bar{\lambda}P_R\lambda, P_R\Lambda,-D_+^2\},\\
&& (V)_D = (D,\cancel{\mathcal{D}}\lambda,0,0,\mathcal{D}^{b}\hat{F}_{ab},-\cancel{\mathcal{D}}\cancel{\mathcal{D}}\lambda,-\square^CD),
\end{eqnarray}
where
\begin{eqnarray}
&& P_L\Lambda \equiv \sqrt{2}P_L(-\frac{1}{2}\gamma\cdot \hat{F} + iD)\lambda,\qquad P_R\Lambda \equiv \sqrt{2}P_R(-\frac{1}{2}\gamma\cdot \hat{F} - iD)\lambda,\\
&& D_-^2 \equiv D^2 - \hat{F}^-\cdot\hat{F}^- - 2  \bar{\lambda}P_L\cancel{\mathcal{D}}\lambda,\qquad D_+^2 \equiv D^2 - \hat{F}^+\cdot\hat{F}^+ - 2  \bar{\lambda}P_R\cancel{\mathcal{D}}\lambda,\\
&& \mathcal{D}_{\mu}\lambda \equiv \bigg(\partial_{\mu}-\frac{3}{2}b_{\mu}+\frac{1}{4}w_{\mu}^{ab}\gamma_{ab}-\frac{3}{2}i\gamma_*\mathcal{A}_{\mu}\bigg)\lambda - \bigg(\frac{1}{4}\gamma^{ab}\hat{F}_{ab}+\frac{1}{2}i\gamma_* D\bigg)\psi_{\mu}
\\
 && \hat{F}_{ab} \equiv F_{ab} + e_a^{~\mu}e_b^{~\nu} \bar{\psi}_{[\mu}\gamma_{\nu]}\lambda,\qquad F_{ab} \equiv e_a^{~\mu}e_b^{~\nu} (2\partial_{[\mu}A_{\nu]}),\\
 && \hat{F}^{\pm}_{\mu\nu} \equiv \frac{1}{2}(\hat{F}_{\mu\nu}\pm \tilde{\hat{F}}_{\mu\nu}), \qquad \tilde{\hat{F}}_{\mu\nu} \equiv -\frac{1}{2} i\epsilon_{\mu\nu\rho\sigma}\hat{F}^{\rho\sigma} .
\end{eqnarray}

Next, we exhibit the components of the superconformal {\it composite} complex multiplets $w'^2$ and $\Bar{w}'^2$ with Weyl/chiral weights $(-1,3)$ and $(-1,-3)$ respectively. These composite multiplets are 
\begin{eqnarray}
&& w'^2 \equiv \frac{\bar{\lambda}P_L\lambda}{(S_0\bar{S}_0e^{-K/3})^2} = \{\mathcal{C}_w,\mathcal{Z}_w,\mathcal{H}_w,\mathcal{K}_w,\mathcal{B}^w_{\mu},\Lambda_w,\mathcal{D}_w\} , \\
&& \bar{w}'^2 \equiv \frac{\bar{\lambda}P_R\lambda}{(S_0\bar{S}_0e^{-K/3})^2}
= \{\mathcal{C}_{\bar{w}},\mathcal{Z}_{\bar{w}},\mathcal{H}_{\bar{w}},\mathcal{K}_{\bar{w}},\mathcal{B}^{\bar{w}}_{\mu},\Lambda_{\bar{w}},\mathcal{D}_{\bar{w}}\}.
\end{eqnarray}
where
\begin{eqnarray}
\mathcal{C}_w &=& h \equiv \frac{\bar{\lambda}P_L\lambda}{(s_0\bar{s}_0e^{-K(z,\bar{z})/3})^2},\\
\mathcal{Z}_w &=& i\sqrt{2}(-h_a\Omega^a + h_{\bar{a}}\Omega^{\bar{a}}),\\
\mathcal{H}_w &=& -2h_aF^a + h_{ab}\bar{\Omega}^a\Omega^b,\\ 
\mathcal{K}_w &=& -2h_{\bar{a}}F^{\bar{a}} + h_{\bar{a}\bar{b}}\bar{\Omega}^{\bar{a}}\Omega^{\bar{b}},\\ 
\mathcal{B}^w_{\mu} &=& ih_a\mathcal{D}_{\mu}X^a-ih_{\bar{a}}\mathcal{D}_{\mu}\bar{X}^{\bar{a}}+ih_{a\bar{b}}\bar{\Omega}^{a}\gamma_{\mu}\Omega^{\bar{b}},\\ 
P_L\Lambda_w &=& -\sqrt{2}ih_{\bar{a}b}[(\cancel{\mathcal{D}}X^b)\Omega^{\bar{a}}-F^{\bar{a}}\Omega^b]-\frac{i}{\sqrt{2}}h_{\bar{a}\bar{b}c}\Omega^c\bar{\Omega}^{\bar{a}}\Omega^{\bar{b}},\\
P_R\Lambda_w &=& \sqrt{2}ih_{a\bar{b}}[(\cancel{\mathcal{D}}\bar{X}^{\bar{b}})\Omega^{a}-F^{a}\Omega^{\bar{b}}]+\frac{i}{\sqrt{2}}h_{ab\bar{c}}\Omega^{\bar{c}}\bar{\Omega}^{a}\Omega^{b},\\
\mathcal{D}_w &=& 2h_{a\bar{b}}\Big(-\mathcal{D}_{\mu}X^a\mathcal{D}^{\mu}\bar{X}^{\bar{b}}-\frac{1}{2}\bar{\Omega}^aP_L\cancel{\mathcal{D}}\Omega^{\bar{b}}-\frac{1}{2}\bar{\Omega}^{\bar{b}}P_R\cancel{\mathcal{D}}\Omega^a+F^aF^{\bar{b}}\Big) \nonumber\\
&&+h_{ab\bar{c}}(-\bar{\Omega}^a\Omega^bF^{\bar{c}}+\bar{\Omega}^a(\cancel{\mathcal{D}}X^b)\Omega^{\bar{c}})+ h_{\bar{a}\bar{b}c}(-\bar{\Omega}^{\bar{a}}\Omega^{\bar{b}}F^{c}+\bar{\Omega}^{\bar{a}}(\cancel{\mathcal{D}}\bar{X}^{\bar{b}})\Omega^{c}) \nonumber\\
&&+ \frac{1}{2}h_{ab\bar{c}\bar{d}}(\bar{\Omega}^aP_L\Omega^b)(\bar{\Omega}^{\bar{c}}P_R\Omega^{\bar{d}}).
\end{eqnarray}
Notice that when finding the multiplet $\bar{w}'^2$, we can just replace $h$ by its complex conjugate $h^{*}$.

The second types of superconformal multiplets that we need are the {\it composite} chiral projection multiplets $T(\Bar{w}'^2)$ and $\Bar{T}(w'^2)$ with Weyl/chiral weights $(0,0)$. From their component supermultiplets defined by 
\begin{eqnarray}
T(\bar{w}'^2) &=& \left( -\frac{1}{2}\mathcal{K}_{\bar{w}}, -\frac{1}{2} \sqrt{2} iP_L (\cancel{\mathcal{D}}\mathcal{Z}_{\bar{w}}+\Lambda_{\bar{w}}), \frac{1}{2}(\mathcal{D}_{\bar{w}}+\square^C \mathcal{C}_{\bar{w}} + i\mathcal{D}_a \mathcal{B}^a_{\bar{w}}) \right),\\
\bar{T}(w'^2) &=& \left( -\frac{1}{2}\mathcal{K}_{\bar{w}}^{*}, \frac{1}{2} \sqrt{2} iP_R (\cancel{\mathcal{D}}\mathcal{Z}_{\bar{w}}^C+\Lambda_{\bar{w}}^C), \frac{1}{2}(\mathcal{D}_{\bar{w}}^{*}+\square^C \mathcal{C}_{\bar{w}}^{*} - i\mathcal{D}_a (\mathcal{B}^a_{\bar{w}})^{*}) \right) ,
\end{eqnarray}
we find the corresponding superconformal multiplets and their complex conjugates as follows:
\begin{eqnarray}
T \equiv T(\bar{w}'^2) &=& \{\mathcal{C}_T,\mathcal{Z}_T,\mathcal{H}_T,\mathcal{K}_T,\mathcal{B}_{\mu}^T,\Lambda_T,\mathcal{D}_T\} ,\nonumber\\
\bar{T} \equiv \bar{T}(w'^2) &=& \{\mathcal{C}_{\bar{T}},\mathcal{Z}_{\bar{T}},\mathcal{H}_{\bar{T}},\mathcal{K}_{\bar{T}},\mathcal{B}_{\mu}^{\bar{T}},\Lambda_{\bar{T}},\mathcal{D}_{\bar{T}}\},
\end{eqnarray}
whose superconformal components are given by
\begin{eqnarray}
\mathcal{C}_T &=&  -\frac{1}{2} \mathcal{K}_{\bar{w}} = h^{*}_{\bar{a}}F^{\bar{a}} -\frac{1}{2} h^{*}_{\bar{a}\bar{b}}\bar{\Omega}^{\bar{a}}\Omega^{\bar{b}} \equiv C_T , \\
\mathcal{Z}_T &=& -\sqrt{2}iP_L\bigg[\cancel{\mathcal{D}}(-h^{*}_a\Omega^a + h^{*}_{\bar{a}}\Omega^{\bar{a}})-h^{*}_{\bar{a}b}[(\cancel{\mathcal{D}}X^b)\Omega^{\bar{a}}-F^{\bar{a}}\Omega^b]-\frac{1}{2}h^{*}_{\bar{a}\bar{b}c}\Omega^c\bar{\Omega}^{\bar{a}}\Omega^{\bar{b}}\nonumber\\
&&\qquad\qquad+h^{*}_{a\bar{b}}[(\cancel{\mathcal{D}}\bar{X}^{\bar{b}})\Omega^{a}-F^{a}\Omega^{\bar{b}}]+\frac{1}{2}h^{*}_{ab\bar{c}}\Omega^{\bar{c}}\bar{\Omega}^{a}\Omega^{b}\bigg] \equiv -\sqrt{2}iP_L\Omega_T ,\\
\mathcal{H}_T &=& -2\bigg[h^{*}_{a\bar{b}}\Big(-\mathcal{D}_{\mu}X^a\mathcal{D}^{\mu}\bar{X}^{\bar{b}}-\frac{1}{2}\bar{\Omega}^aP_L\cancel{\mathcal{D}}\Omega^{\bar{b}}-\frac{1}{2}\bar{\Omega}^{\bar{b}}P_R\cancel{\mathcal{D}}\Omega^a+F^aF^{\bar{b}}\Big) \nonumber\\
&&+\frac{1}{2}h^{*}_{ab\bar{c}}(-\bar{\Omega}^a\Omega^bF^{\bar{c}}+\bar{\Omega}^a(\cancel{\mathcal{D}}X^b)\Omega^{\bar{c}})+\frac{1}{2} h^{*}_{\bar{a}\bar{b}c}(-\bar{\Omega}^{\bar{a}}\Omega^{\bar{b}}F^{c}+\bar{\Omega}^{\bar{a}}(\cancel{\mathcal{D}}\bar{X}^{\bar{b}})\Omega^{c}) \nonumber\\
&&+ \frac{1}{4}h^{*}_{ab\bar{c}\bar{d}}(\bar{\Omega}^aP_L\Omega^b)(\bar{\Omega}^{\bar{c}}P_R\Omega^{\bar{d}})+\frac{1}{2}\square^C h^{*} + \frac{1}{2}i\mathcal{D}^{\mu} (ih^{*}_a\mathcal{D}_{\mu}X^a-ih^{*}_{\bar{a}}\mathcal{D}_{\mu}\bar{X}^{\bar{a}}+ih^{*}_{a\bar{b}}\bar{\Omega}^{a}\gamma_{\mu}\Omega^{\bar{b}})\bigg] \nonumber\\
&\equiv& -2F_T,\\
\mathcal{K}_T &=& 0,\\
\mathcal{B}^T_{\mu} &=& -i\mathcal{D}_{\mu}\mathcal{C}_T,\\
\Lambda_T &=& 0 ,\\
\mathcal{D}_T &=& 0,
\end{eqnarray}
where we used $a,b,c,d = 0,i(\equiv z^i),W (\equiv \bar{\lambda}P_L\lambda)$. This gives the superfield components of the chiral projection multiplet $T$:
\begin{eqnarray}
T(\bar{w}'^2) = ( C_T, P_L\Omega_T, F_T )
\end{eqnarray}
where 
\begin{eqnarray}
C_T &=&  h^{*}_{\bar{a}}F^{\bar{a}} -\frac{1}{2} h^{*}_{\bar{a}\bar{b}}\bar{\Omega}^{\bar{a}}\Omega^{\bar{b}},\\
P_L\Omega_T &=& \cancel{\mathcal{D}}(-h^{*}_a\Omega^a + h^{*}_{\bar{a}}\Omega^{\bar{a}})-h^{*}_{\bar{a}b}[(\cancel{\mathcal{D}}X^b)\Omega^{\bar{a}}-F^{\bar{a}}\Omega^b]-\frac{1}{2}h^{*}_{\bar{a}\bar{b}c}\Omega^c\bar{\Omega}^{\bar{a}}\Omega^{\bar{b}}\nonumber\\
&&+h^{*}_{a\bar{b}}[(\cancel{\mathcal{D}}\bar{X}^{\bar{b}})\Omega^{a}-F^{a}\Omega^{\bar{b}}]+\frac{1}{2}h^{*}_{ab\bar{c}}\Omega^{\bar{c}}\bar{\Omega}^{a}\Omega^{b},\\
F_T &=&  h^{*}_{a\bar{b}}\Big(-\mathcal{D}_{\mu}X^a\mathcal{D}^{\mu}\bar{X}^{\bar{b}}-\frac{1}{2}\bar{\Omega}^aP_L\cancel{\mathcal{D}}\Omega^{\bar{b}}-\frac{1}{2}\bar{\Omega}^{\bar{b}}P_R\cancel{\mathcal{D}}\Omega^a+F^aF^{\bar{b}}\Big) \nonumber\\
&&+\frac{1}{2}h^{*}_{ab\bar{c}}(-\bar{\Omega}^a\Omega^bF^{\bar{c}}+\bar{\Omega}^a(\cancel{\mathcal{D}}X^b)\Omega^{\bar{c}})+\frac{1}{2} h^{*}_{\bar{a}\bar{b}c}(-\bar{\Omega}^{\bar{a}}\Omega^{\bar{b}}F^{c}+\bar{\Omega}^{\bar{a}}(\cancel{\mathcal{D}}\bar{X}^{\bar{b}})\Omega^{c}) \nonumber\\
&&+ \frac{1}{4}h^{*}_{ab\bar{c}\bar{d}}(\bar{\Omega}^aP_L\Omega^b)(\bar{\Omega}^{\bar{c}}P_R\Omega^{\bar{d}})+\frac{1}{2}\square^C h^{*} - \frac{1}{2}\mathcal{D}^{\mu} (h^{*}_a\mathcal{D}_{\mu}X^a-h^{*}_{\bar{a}}\mathcal{D}_{\mu}\bar{X}^{\bar{a}}+h^{*}_{a\bar{b}}\bar{\Omega}^{a}\gamma_{\mu}\Omega^{\bar{b}}). \nonumber\\{}
\end{eqnarray}

Morever,
\begin{eqnarray}
\bar{T}(w'^2) = \{ C_T^{*}, P_R\Omega_T, F_T^{*}\}
\end{eqnarray}
where
\begin{eqnarray}
C_T^{*} &=&  h_{a}F^{a} -\frac{1}{2} h_{ab}\bar{\Omega}^{a}\Omega^{b},\\
P_R\Omega_T &=& \cancel{\mathcal{D}}(-h_a\Omega^a + h_{\bar{a}}\Omega^{\bar{a}})-h_{\bar{a}b}[(\cancel{\mathcal{D}}X^b)\Omega^{\bar{a}}-F^{\bar{a}}\Omega^b]-\frac{1}{2}h_{\bar{a}\bar{b}c}\Omega^c\bar{\Omega}^{\bar{a}}\Omega^{\bar{b}}\nonumber\\
&&+h_{a\bar{b}}[(\cancel{\mathcal{D}}\bar{X}^{\bar{b}})\Omega^{a}-F^{a}\Omega^{\bar{b}}]+\frac{1}{2}h_{ab\bar{c}}\Omega^{\bar{c}}\bar{\Omega}^{a}\Omega^{b},\\
F_T^{*} &=&  h_{a\bar{b}}\Big(-\mathcal{D}_{\mu}X^a\mathcal{D}^{\mu}\bar{X}^{\bar{b}}-\frac{1}{2}\bar{\Omega}^aP_L\cancel{\mathcal{D}}\Omega^{\bar{b}}-\frac{1}{2}\bar{\Omega}^{\bar{b}}P_R\cancel{\mathcal{D}}\Omega^a+F^aF^{\bar{b}}\Big) \nonumber\\
&&+\frac{1}{2}h_{ab\bar{c}}(-\bar{\Omega}^a\Omega^bF^{\bar{c}}+\bar{\Omega}^a(\cancel{\mathcal{D}}X^b)\Omega^{\bar{c}})+\frac{1}{2} h_{\bar{a}\bar{b}c}(-\bar{\Omega}^{\bar{a}}\Omega^{\bar{b}}F^{c}+\bar{\Omega}^{\bar{a}}(\cancel{\mathcal{D}}\bar{X}^{\bar{b}})\Omega^{c}) \nonumber\\
&&+ \frac{1}{4}h_{ab\bar{c}\bar{d}}(\bar{\Omega}^aP_L\Omega^b)(\bar{\Omega}^{\bar{c}}P_R\Omega^{\bar{d}})+\frac{1}{2}\square^C h - \frac{1}{2}\mathcal{D}^{\mu} (h_a\mathcal{D}_{\mu}X^a-h_{\bar{a}}\mathcal{D}_{\mu}\bar{X}^{\bar{a}}+h_{a\bar{b}}\bar{\Omega}^{\bar{b}}\gamma_{\mu}\Omega^{a}). \nonumber\\{}
\end{eqnarray}

We then present a superconformal composite real multiplet $\mathcal{R}$ with Weyl/chiral weights $(0,0)$. By introducing the chiral multiplets $\mathcal{X}^A \equiv \{X^A,P_L\Omega^A,F^A\}$ where $A=\{ S_0,Z^i,\bar{\lambda}P_L\lambda,T(\bar{w}'^2)$\} and their conjugates, we represent  $\mathcal{R}$ as
\begin{eqnarray}
\mathcal{R} \equiv (S_0\bar{S}_0e^{-K/3})^{-3} \frac{(\bar{\lambda}P_L\lambda)(\bar{\lambda}P_R\lambda)}{T(\bar{w}'^2) \bar{T}(w'^2)} \mathcal{U}
\end{eqnarray}
whose lowest component is 
\begin{eqnarray}
\mathcal{C}_{\mathcal{R}}  \equiv (s_0\bar{s}_0e^{-K/3})^{-3}\frac{(\bar{\lambda}P_L\lambda)(\bar{\lambda}P_R\lambda)}{C_TC_{\bar{T}}}\mathcal{U} 
\equiv f(X^A,\bar{X}^{\bar{A}}) \label{def-f}
\end{eqnarray}
where $C_T = -D_+^2 \Delta^{-2}$; $C_{\bar{T}} = -D_-^2\Delta^{-2}$, and $\Delta \equiv s_0\bar{s}_0e^{-K/3}$, and $K,\mathcal{U}$ are functions of the matter multiplets $Z^i$'s.

\begin{eqnarray}
\mathcal{C}_{\mathcal{R}} &=& f \equiv  (s_0\bar{s}_0e^{-K/3})^{-3}\frac{(\bar{\lambda}P_L\lambda)(\bar{\lambda}P_R\lambda)}{C_TC_{\bar{T}}}\mathcal{U} ,\\
\mathcal{Z}_{\mathcal{R}} &=& i\sqrt{2}(-f_A\Omega^A + f_{\bar{A}}\Omega^{\bar{A}}),\\
\mathcal{H}_{\mathcal{R}} &=& -2f_AF^A + f_{AB}\bar{\Omega}^A\Omega^B,\\ 
\mathcal{K}_{\mathcal{R}} &=& -2f_{\bar{A}}F^{\bar{A}} + f_{\bar{A}\bar{B}}\bar{\Omega}^{\bar{A}}\Omega^{\bar{B}},\\ 
\mathcal{B}^{\mathcal{R}}_{\mu} &=& if_A\mathcal{D}_{\mu}X^A-if_{\bar{A}}\mathcal{D}_{\mu}\bar{X}^{\bar{A}}+if_{A\bar{B}}\bar{\Omega}^{A}\gamma_{\mu}\Omega^{\bar{B}},\\ 
P_L\Lambda_{\mathcal{R}} &=& -\sqrt{2}if_{\bar{A}B}[(\cancel{\mathcal{D}}X^B)\Omega^{\bar{A}}-F^{\bar{A}}\Omega^B]-\frac{i}{\sqrt{2}}f_{\bar{A}\bar{B}C}\Omega^C\bar{\Omega}^{\bar{A}}\Omega^{\bar{B}},\\
P_R\Lambda_{\mathcal{R}} &=& \sqrt{2}if_{A\bar{B}}[(\cancel{\mathcal{D}}\bar{X}^{\bar{B}})\Omega^{A}-F^{A}\Omega^{\bar{B}}]+\frac{i}{\sqrt{2}}f_{AB\bar{C}}\Omega^{\bar{C}}\bar{\Omega}^{A}\Omega^{B},\\
\mathcal{D}_{\mathcal{R}} &=& 2f_{A\bar{B}}\Big(-\mathcal{D}_{\mu}X^A\mathcal{D}^{\mu}\bar{X}^{\bar{B}}-\frac{1}{2}\bar{\Omega}^AP_L\cancel{\mathcal{D}}\Omega^{\bar{B}}-\frac{1}{2}\bar{\Omega}^{\bar{B}}P_R\cancel{\mathcal{D}}\Omega^A+F^AF^{\bar{B}}\Big) \nonumber\\
&&+f_{AB\bar{C}}(-\bar{\Omega}^A\Omega^BF^{\bar{C}}+\bar{\Omega}^A(\cancel{\mathcal{D}}X^B)\Omega^{\bar{C}})+ f_{\bar{A}\bar{B}C}(-\bar{\Omega}^{\bar{A}}\Omega^{\bar{B}}F^{C}+\bar{\Omega}^{\bar{A}}(\cancel{\mathcal{D}}\bar{X}^{\bar{B}})\Omega^{C}) \nonumber\\
&&+ \frac{1}{2}f_{AB\bar{C}\bar{D}}(\bar{\Omega}^AP_L\Omega^B)(\bar{\Omega}^{\bar{C}}P_R\Omega^{\bar{D}}).
\end{eqnarray}

Then, the superconformal multiplet of the new Fayet-Iliopoulos term can be written by using
\begin{eqnarray}
\mathcal{R}\cdot (V)_D = \{\tilde{\mathcal{C}},\tilde{\mathcal{Z}},\tilde{\mathcal{H}},\tilde{\mathcal{K}},\tilde{\mathcal{B}}_{\mu},\tilde{\Lambda},\tilde{\mathcal{D}}\},
\end{eqnarray}
whose superconformal multiplet components are as follows:
\begin{eqnarray}
\tilde{\mathcal{C}} &=& Df,\\
\tilde{\mathcal{Z}} &=& f\cancel{\mathcal{D}}\lambda+Di\sqrt{2}(-f_{A}\Omega^A+f_{\bar{A}}\Omega^{\bar{A}}),\\
\tilde{\mathcal{H}} &=& D(-2f_AF^A + f_{AB}\bar{\Omega}^A\Omega^B)-i\sqrt{2}(-f_{A}\bar{\Omega}^A+f_{\bar{A}}\bar{\Omega}^{\bar{A}})P_L\cancel{\mathcal{D}}\lambda,\\
\tilde{\mathcal{K}} &=& D(-2f_{\bar{A}}F^{\bar{A}} + f_{\bar{A}\bar{B}}\bar{\Omega}^{\bar{A}}\Omega^{\bar{B}})-i\sqrt{2}(-f_{A}\bar{\Omega}^A+f_{\bar{A}}\bar{\Omega}^{\bar{A}})P_R\cancel{\mathcal{D}}\lambda,\\
\tilde{\mathcal{B}} &=& (\mathcal{D}^{\nu}\hat{F}_{\mu\nu})f+D(if_A\mathcal{D}_{\mu}X^A-if_{\bar{A}}\mathcal{D}_{\mu}\bar{X}^{\bar{A}}+if_{A\bar{B}}\bar{\Omega}^A\gamma_{\mu}\Omega^{\bar{B}}),\\
\tilde{\Lambda} &=& -f\cancel{\mathcal{D}}\cancel{\mathcal{D}}\lambda + D(P_L\Lambda_{\mathcal{R}}+P_R\Lambda_{\mathcal{R}})+\frac{1}{2}\Big(\gamma_*(-f_A\cancel{\mathcal{D}}X^A+f_{\bar{A}}\cancel{\mathcal{D}}\bar{X}^{\bar{A}}-f_{A\bar{B}}\bar{\Omega}^A\cancel{\gamma}\Omega^{\bar{B}})\nonumber\\
&&+P_L(-2f_{\bar{A}}F^{\bar{A}} + f_{\bar{A}\bar{B}}\bar{\Omega}^{\bar{A}}\Omega^{\bar{B}})+P_R(-2f_AF^A + f_{AB}\bar{\Omega}^A\Omega^B) -\cancel{\mathcal{D}}f\Big)\cancel{\mathcal{D}}\lambda\nonumber\\
&& +\frac{1}{2}\Big( i\gamma_*\gamma^{\mu}\mathcal{D}^{\nu}\hat{F}_{\mu\nu}   -\cancel{\mathcal{D}}D\Big)i\sqrt{2}(-f_{A}\Omega^A+f_{\bar{A}}\Omega^{\bar{A}}) ,\\
\tilde{\mathcal{D}} &=&-f\square^C D + D 
\bigg\{ 2f_{A\bar{B}}(-\mathcal{D}_{\mu}X^A\mathcal{D}^{\mu}\bar{X}^{\bar{B}}-\frac{1}{2}\bar{\Omega}^AP_L\cancel{\mathcal{D}}\Omega^{\bar{B}}-\frac{1}{2}\cancel{\Omega}^{\bar{B}}P_R\cancel{\mathcal{D}}\Omega^A+F^AF^{\bar{B}}) \nonumber\\
&&+f_{AB\bar{C}}(-\bar{\Omega}^A\Omega^B F^{\bar{C}} + \bar{\Omega}^A(\cancel{\mathcal{D}}X^B)\Omega^{\bar{C}}) 
+f_{\bar{A}\bar{B}C}(-\bar{\Omega}^{\bar{A}}\Omega^{\bar{B}} F^C + \bar{\Omega}^{\bar{A}}(\cancel{\mathcal{D}}\bar{X}^{\bar{B}})\Omega^C) \nonumber\\
&&+\frac{1}{2}f_{AB\bar{C}\bar{D}} (\bar{\Omega}^AP_L\Omega^B)(\bar{\Omega}^{\bar{C}}P_R\Omega^{\bar{D}}) \bigg\}\nonumber\\
&& -(\mathcal{D}_{\nu}\hat{F}^{\mu\nu})(if_A\mathcal{D}_{\mu}X^A-if_{\bar{A}}\mathcal{D}_{\mu}\bar{X}^{\bar{A}}+if_{A\bar{B}}\bar{\Omega}^A\gamma_{\mu}\Omega^{\bar{B}}) \nonumber\\
&& + \bigg( \sqrt{2}if_{\bar{A}B}[(\cancel{\mathcal{D}}X^B)\Omega^{\bar{A}}-F^{\bar{A}}\Omega^B]  +\frac{i}{\sqrt{2}}f_{\bar{A}\bar{B}C} \Omega^C\bar{\Omega}^{\bar{A}}\Omega^{\bar{B}} \bigg)\cancel{\mathcal{D}}\lambda\nonumber\\
&& - \bigg( \sqrt{2}if_{A\bar{B}}[(\cancel{\mathcal{D}}\bar{X}^{\bar{B}})\Omega^{A}-F^{A}\Omega^{\bar{B}}]  +\frac{i}{\sqrt{2}}f_{AB\bar{C}} \Omega^{\bar{C}}\bar{\Omega}^{A}\Omega^{B} \bigg)\cancel{\mathcal{D}}\lambda\nonumber\\
&&-(\mathcal{D}_{\mu}f)(\mathcal{D}^{\mu}D)-\frac{1}{2}\cancel{\mathcal{D}}[i\sqrt{2}(-f_{A}\Omega^A+f_{\bar{A}}\Omega^{\bar{A}})](\cancel{\mathcal{D}}\lambda)+\frac{1}{2}i\sqrt{2}(-f_{A}\Omega^A+f_{\bar{A}}\Omega^{\bar{A}})(\cancel{\mathcal{D}}\cancel{\mathcal{D}}\lambda),\nonumber \\ &&
\end{eqnarray}
where the indices $A,B,C,D$ run over $0,i,W,T$. The component action of the new FI term is then given by the D-term density formula 
\begin{eqnarray}
\mathcal{L}_{NEW} \equiv -[ \mathcal{R}\cdot (V)_D]_D &=& -\frac{1}{4}\int d^4x e \bigg[  \tilde{\mathcal{D}} -\frac{1}{2}\bar{\psi}\cdot \gamma i\gamma_* \tilde{\Lambda} -\frac{1}{3}\tilde{\mathcal{C}}R(\omega)\nonumber\\
&&+\frac{1}{6}\Big(\tilde{\mathcal{C}}\bar{\psi}_{\mu}\gamma^{\mu\rho\sigma}-i\bar{\tilde{\mathcal{Z}}}\gamma^{\rho\sigma}\gamma_*\Big)R'_{\rho\sigma}(Q)\nonumber\\
&&+\frac{1}{4}\varepsilon^{abcd}\bar{\psi}_{a}\gamma_b\psi_c\Big(\tilde{\mathcal{B}}_{d}-\frac{1}{2}\bar{\psi}_d\tilde{\mathcal{Z}}\Big)\bigg]+\textrm{h.c.}.
\end{eqnarray}

Using $f=\Delta^{-3}\frac{W\bar{W}}{C_TC_{\bar{T}}}\mathcal{U} $, $W\equiv(\bar{\lambda}P_L\lambda)$, $\bar{W}\equiv (\bar{\lambda}P_R\lambda)$, $\Omega^W \sim \sqrt{2}iDP_L\lambda$, $\bar{\Omega}^T \sim 2D^2\Delta^{-2}(\frac{\bar{\Omega}^0}{s_0}-\frac{K_I\bar{\Omega}^I}{3}) $, $C_T \sim -D^2\Delta^{-2}$, $F^W \sim -D^2$, $F^{\bar{T}} \sim 2D^2\Delta^{-2}( \frac{F^{\bar{0}}}{\bar{s}_0} - \frac{1}{3}K_{\bar{J}}F^{\bar{J}})$ where $\Delta \equiv s_0\bar{s}_0e^{-K/3}$,
\begin{eqnarray}
&& \mathcal{L}_{\textrm{newFI}}^{(\textrm{2f})}e^{-1} \nonumber\\
&&= -D f_{0\bar{W}}F^0F^{\bar{W}}  -D f_{I\bar{W}}F^IF^{\bar{W}} -D f_{T\bar{W}}F^TF^{\bar{W}} \nonumber\\
&&\quad + \frac{1}{2}f_{0W\bar{W}}\bar{\Omega}^0\Omega^WF^{\bar{W}}
+ \frac{1}{2}f_{IW\bar{W}}\bar{\Omega}^I\Omega^WF^{\bar{W}} 
+ \frac{1}{2}f_{TW\bar{W}}\bar{\Omega}^T\Omega^WF^{\bar{W}} 
- \frac{D\sqrt{2}}{4} \bar{\psi}_{\mu} \gamma^{\mu} f_{\bar{W}W}F^{\bar{W}}\Omega^W 
+c.c..,\nonumber\\
&&=  -3\Delta\frac{(\bar{\lambda}P_L\lambda)}{D}\mathcal{U}\frac{F^{0}}{s_0} +\Delta\frac{(\bar{\lambda}P_L\lambda)}{D}(K_I\mathcal{U}+\mathcal{U}_I)F^{I}  +2\Delta\frac{(\bar{\lambda}P_L\lambda)}{D}\mathcal{U}( \frac{F^0}{s_0} - \frac{1}{3}K_{I}F^{I})\nonumber\\
&&\quad + \frac{3i}{\sqrt{2}}\frac{\Delta}{Ds_0}\mathcal{U}(\bar{\Omega}^0P_L\lambda) - \frac{i}{\sqrt{2}}\frac{\Delta}{D}(K_I\mathcal{U}+\mathcal{U}_I)(\bar{\Omega}^IP_L\lambda) + \sqrt{2}i \frac{\Delta}{Ds_0}\mathcal{U}(\bar{\Omega}^0P_L\lambda)
-\frac{\sqrt{2}i}{3} \frac{\Delta}{D}\mathcal{U}K_I(\bar{\Omega}^IP_L\lambda)\nonumber\\
&&\quad + \frac{i}{2}\Delta \mathcal{U} (\bar{\psi}_{\mu} \gamma^{\mu}P_L\lambda) + c.c.,\nonumber\\
&&= \frac{\Delta}{D}\bigg(
-\frac{F^0\mathcal{U}}{s_0}  + \mathcal{U}_IF^I +\frac{1}{3}\mathcal{U}K_IF^I
\bigg)(\bar{\lambda}P_L\lambda) 
+\frac{5}{\sqrt{2}}i \frac{\Delta}{Ds_0}\mathcal{U}(\bar{\Omega}^0P_L\lambda) 
\nonumber\\
&&\quad -i \frac{\Delta}{D}  \bigg(
\frac{5}{3\sqrt{2}} K_I \mathcal{U} +\frac{1}{\sqrt{2}}\mathcal{U}_I
\bigg)(\bar{\Omega}^IP_L\lambda) + \frac{i}{2}\Delta \mathcal{U} (\bar{\psi}_{\mu} \gamma^{\mu}P_L\lambda) + h.c.
\end{eqnarray}

In the superconformal gauge (i.e. $P_L\Omega^0 = \frac{1}{3}e^{K/6}K_IP_L\Omega^I$, $s_0=\bar{s}_0=e^{K/6}$, $\Delta = 1$), the Lagrangian   is
\begin{eqnarray}
&&\mathcal{L}_{\textrm{newFI}}^{(\textrm{2f})}e^{-1} \nonumber\\
&&= \frac{1}{D}\bigg(
-F^0\mathcal{U}e^{-K/6}  + \mathcal{U}_IF^I +\frac{1}{3}\mathcal{U}K_IF^I
\bigg)(\bar{\lambda}P_L\lambda) 
 -\frac{i}{D}  
\frac{\mathcal{U}_I}{\sqrt{2}}(\bar{\Omega}^IP_L\lambda) + \frac{i}{2} \mathcal{U} (\bar{\psi}_{\mu} \gamma^{\mu}P_L\lambda) + h.c. \nonumber\\
\end{eqnarray}

The D-term Lagrangian   is found to be
\begin{eqnarray}
\mathcal{L}_De^{-1} &\supset& \frac{1}{2}D^2 - \mathcal{U}D + \frac{1}{D}\bigg(
-F^0\mathcal{U}e^{-K/6}  + \mathcal{U}_IF^I +\frac{1}{3}\mathcal{U}K_IF^I
\bigg)(\bar{\lambda}P_L\lambda) 
 -\frac{i}{D}  
\frac{\mathcal{U}_I}{\sqrt{2}}(\bar{\Omega}^IP_L\lambda)\nonumber\\
&& + \frac{1}{D}\bigg(
-F^{\bar{0}}\mathcal{U}e^{-K/6}  + \mathcal{U}_{\bar{J}}F^{\bar{J}} +\frac{1}{3}\mathcal{U}K_{\bar{J}}F^{\bar{J}}
\bigg)(\bar{\lambda}P_R\lambda) 
 +\frac{i}{D}  
\frac{\mathcal{U}_{\bar{J}}}{\sqrt{2}}(\bar{\Omega}^{\bar{J}}P_R\lambda) .
\end{eqnarray}
The solution for $D$ is
\begin{eqnarray}
D= \mathcal{U} + \frac{1}{\mathcal{U}^2}
\bigg[
\bigg(
-F^0\mathcal{U}e^{-K/6}  + \mathcal{U}_IF^I +\frac{1}{3}\mathcal{U}K_IF^I
\bigg)(\bar{\lambda}P_L\lambda)  -i 
\frac{\mathcal{U}_I}{\sqrt{2}}(\bar{\Omega}^IP_L\lambda) + h.c.
\bigg]+\textrm{higher order terms},\nonumber\\{}
\end{eqnarray}
Then, we find
\begin{eqnarray}
\mathcal{L}_{\textrm{newFI}}^{(2f)}e^{-1} = \frac{1}{\mathcal{U}}
\bigg[
\bigg(
-F^0\mathcal{U}e^{-K/6}  + \mathcal{U}_IF^I +\frac{1}{3}\mathcal{U}K_IF^I
\bigg)(\bar{\lambda}P_L\lambda)  -i 
\frac{\mathcal{U}_I}{\sqrt{2}}(\bar{\Omega}^IP_L\lambda) + h.c.
\bigg] .
\end{eqnarray}
The total Lagrangian   containing the auxiliary fields $F^0$ and $F^I$ is given by
\begin{eqnarray}
\mathcal{L}e^{-1} &=& -3e^{-K/3}F^0F^{\bar{0}} + 3e^{K/3}WF^0 + 3e^{K/3}\bar{W}F^{\bar{0}} + \frac{1}{9}G_{I\bar{J}}F^IF^{\bar{J}} \nonumber\\
&& + \frac{1}{3}e^{K/2}\nabla_IWF^I + \frac{1}{3}e^{K/2}\nabla_{\bar{J}}\bar{W}F^{\bar{J}}\nonumber\\
&& + \frac{1}{\mathcal{U}}
\bigg[
\bigg(
-F^0\mathcal{U}e^{-K/6}  + \mathcal{U}_IF^I +\frac{1}{3}\mathcal{U}K_IF^I
\bigg)(\bar{\lambda}P_L\lambda)  -i 
\frac{\mathcal{U}_I}{\sqrt{2}}(\bar{\Omega}^IP_L\lambda) 
\bigg]
\nonumber\\
&& + \frac{1}{\mathcal{U}}
\bigg[
\bigg(
-F^{\bar{0}}\mathcal{U}e^{-K/6}  + \mathcal{U}_{\bar{J}}F^{\bar{J}} +\frac{1}{3}\mathcal{U}K_{\bar{J}}F^{\bar{J}}
\bigg)(\bar{\lambda}P_R\lambda)  +i 
\frac{\mathcal{U}_{\bar{J}}}{\sqrt{2}}(\bar{\Omega}^{\bar{J}}P_R\lambda)
\bigg],
\end{eqnarray}
where $\nabla_IW \equiv W_I + K_I W$. By solving the equations of motion for the auxiliary fields, we find 
\begin{eqnarray}
F^0 &=& e^{2K/3}\bar{W} - \frac{1}{3}e^{K/6}(\bar{\lambda}P_R\lambda),\\
F^{\bar{J}} &=&- 3e^{K/2}G^{I\bar{J}}\nabla_IW -G^{I\bar{J}}\Big(9\frac{\mathcal{U}_I }{\mathcal{U}}+ 3K_I\Big)(\bar{\lambda}P_L\lambda)
\end{eqnarray}
and also read off the mass $m_{I\lambda}$
\begin{eqnarray}
m_{I\lambda}^{FI} &=& -\frac{i}{\sqrt{2}}\frac{\mathcal{U}_I}{\mathcal{U}},\\
m_{\lambda\lambda}^{FI} &=& -e^{K/2} \left( \bar{W} + 4G^{I\bar{J}}\left(\frac{\mathcal{U}_I}{\mathcal{U}}+\frac{K_I}{3}\right)(\bar{W}_{\bar{J}}+K_{\bar{J}}\bar{W}) \right) .
\end{eqnarray}

The gravitino mixing term is given by
\begin{eqnarray}
\mathcal{L}_{\textrm{mix}} e^{-1} =
\frac{1}{\sqrt{2}}\nabla_IW e^{K/2} \bar{\psi}_{\mu} \gamma^{\mu} P_L\Omega^I + \frac{i}{2} \mathcal{P}_A \bar{\psi}_{\mu} \gamma^{\mu} P_L\lambda^A + \frac{i}{2}\mathcal{U}\bar{\psi}_{\mu} \gamma^{\mu} P_L\lambda + h.c = -\bar{\psi}_{\mu}\gamma^{\mu} P_Lv+h.c., \nonumber \\ 
\end{eqnarray}
which gives the goldstino 
\begin{eqnarray}
P_Lv = -\frac{1}{\sqrt{2}}\nabla_IW e^{K/2}  P_L\Omega^I - \frac{i}{2} \mathcal{P}_A P_L\lambda^A -\frac{i}{2}\mathcal{U} P_L\lambda ,
\end{eqnarray}
where $\lambda^A$ is the gaugino corresponding to the gauge multiplet $V_A$, and $\lambda$ is the superpartner of the new FI term vector multiplet $V$.

The fermionic masses from standard $\mathcal{N}=1$ supergravity are found to be
\begin{eqnarray}
m_{3/2} &=& e^{K/2}W,\\
m_{IJ}^{(0)} &=& e^{K/2}(\partial_I+K_I)(W_J+K_JW)-e^{K/2} G^{K\bar{L}}\partial_IG_{J\bar{L}}(W_K+K_KW),\\
m_{IA}^{(0)} &=& i\sqrt{2}[\partial_I\mathcal{P}_A-\frac{1}{4}f_{ABI}(\textrm{Re}f)^{-1~BC}\mathcal{P}_C],\\
m_{AB}^{(0)} &=& -\frac{1}{2}e^{K/2}f_{ABI} G^{I\bar{J}}(\bar{W}_{\bar{J}}+K_{\bar{J}}\bar{W}),\\
m_{I\lambda}^{(0)} &=& 0,\\
m_{\lambda\lambda}^{(0)} &=& 0.
\end{eqnarray}
The fermionic masses generated by super-Higgs effect are given by
\begin{eqnarray}
m_{IJ}^{(\nu)} &=& -\frac{2}{3W} e^{K/2} (W_I+K_IW)(W_J+K_JW),\\
m_{IA}^{(\nu)} &=& -i\frac{2}{3\sqrt{2}W}(W_I+K_IW) \mathcal{P}_A ,\\
m_{AB}^{(\nu)} &=& \frac{1}{3e^{K/2}W} \mathcal{P}_A\mathcal{P}_B ,\\
m_{I\lambda}^{(\nu)} &=& -i\frac{2}{3\sqrt{2}W}(W_I+K_IW) \mathcal{U},\\
m_{\lambda\lambda}^{(\nu)} &=& \frac{\mathcal{U}^2}{3e^{K/2}W} .
\end{eqnarray}
The fermionic masses from the new FI term are 
\begin{eqnarray}
 m_{IJ}^{FI} &=& 0 ,\\
m_{IA}^{FI} &=&  0,\\
m_{AB}^{FI} &=&  0,\\
m_{I\lambda}^{FI} &=& -\frac{i}{\sqrt{2}}\frac{\mathcal{U}_I}{\mathcal{U}},\\
m_{\lambda\lambda}^{FI} &=& -e^{K/2} \left( \bar{W} + 4G^{I\bar{J}}\left(\frac{\mathcal{U}_I}{\mathcal{U}}+\frac{K_I}{3}\right)(\bar{W}_{\bar{J}}+K_{\bar{J}}\bar{W}) \right).
\end{eqnarray}
Thus, the final fermionic masses are obtained by combining the three contributions above as follows:
\begin{eqnarray}
 m_{3/2} &=& We^{K/2},\nonumber\\
  m_{IJ}^{(g)} &=& m_{IJ}^{(0)}+m_{IJ}^{FI}+m_{IJ}^{(\nu)} \nonumber\\
&=& e^{K/2}(W_{IJ} + K_{IJ}W+K_JW_I+K_IW_J + K_IK_JW)
\nonumber\\
&& -e^{K/2} G^{K\bar{L}}\partial_I G_{J\bar{L}}(W_K + K_KW) -\frac{2}{3}  (W_I+K_IW)(W_J+K_JW),\nonumber\\
m_{IA}^{(g)} &=& m_{IA}^{(0)}+m_{IA}^{FI}+m_{IA}^{(\nu)} \nonumber\\
&=&
 i\sqrt{2}[\partial_I\mathcal{P}_A-\frac{1}{4}f_{ABI}(\textrm{Re}f)^{-1~BC}\mathcal{P}_C]-i\frac{2}{3\sqrt{2}W}(W_I+K_IW) \mathcal{P}_A\nonumber
\\
m_{AB}^{(g)} &=&m_{IA}^{(0)}+m_{IA}^{FI}+m_{IA}^{(\nu)}  \nonumber\\
&=&  -\frac{1}{2}e^{K/2}f_{ABI} G^{I\bar{J}}(\bar{W}_{\bar{J}}+K_{\bar{J}}\bar{W}) + \frac{1}{3e^{K/2}W} \mathcal{P}_A\mathcal{P}_B\nonumber\\
m_{I\lambda}^{(g)} &=& m_{I\lambda}^{(0)}+m_{I\lambda}^{FI}+m_{I\lambda}^{(\nu)} \nonumber\\
&=& -\frac{i}{\sqrt{2}}\frac{\mathcal{U}_I}{\mathcal{U}}-\frac{i\sqrt{2}}{3W}(W_I+K_IW) \mathcal{U} = m_{\lambda I}^{(g)},\nonumber\\
m_{\lambda\lambda}^{(g)} &=& m_{\lambda\lambda}^{(0)}+m_{\lambda\lambda}^{FI}+m_{\lambda\lambda}^{(\nu)} \nonumber\\
&=& -e^{K/2} \left( \bar{W} + 4G^{I\bar{J}}\left(\frac{\mathcal{U}_I}{\mathcal{U}}+\frac{K_I}{3}\right)(\bar{W}_{\bar{J}}+K_{\bar{J}}\bar{W}) \right) +\frac{\mathcal{U}^2}{3e^{K/2}W}. \label{General_Fermion_Masses}
\end{eqnarray}


\begin{thebibliography}{99}
\bibitem{nonlin} 
Z.~Komargodski and N.~Seiberg, ``From Linear SUSY to Constrained Superfields,''
JHEP \textbf{09}, 066 (2009) [arXiv:0907.2441 [hep-th]].

 \bibitem{cfgvnv} E.~Cremmer, B.~Julia, J.~Scherk, S.~Ferrara, L.~Girardello and P.~van Nieuwenhuizen,
``Spontaneous Symmetry Breaking and Higgs Effect in Supergravity Without Cosmological Constant,''
Nucl. Phys. B \textbf{147}, 105 (1979); E.~Cremmer, S.~Ferrara, L.~Girardello and A.~Van Proeyen,
``Yang-Mills Theories with Local Supersymmetry: Lagrangian  , Transformation Laws and SuperHiggs Effect,''
Nucl. Phys. B \textbf{212}, 413 (1983).


\bibitem{AKK} Y. Aldabergenov, S. V. Ketov, and R. Knoops, ``General couplings of a vector multiplet in $\mathcal{N}=1$ supergravity with new FI terms,'' Phys. Lett. B \textbf{785}, 284 (2018) [arXiv:1806.04290 [hep-th]].

\bibitem{acik} I. Antoniadis, A. Chatrabhuti, H. Isono, and R. Knoops, ``Fayet-Iliopoulos terms in supergravity and D-term inflation'', Eur. Phys. J. C (2018) 78:366 [arXiv:1803.03817 [hep-th]]; I. Antoniadis, A. Chatrabhuti, H. Isono, and R. Knoops, ``The cosmological constant in supergravity,'' Eur. Phys. J. C  \textbf{78}, 718 (2018) [arXiv:1805.00852 [hep-th]].




\bibitem{cftp} N. Cribiori, F. Farakos, M. Tournoy, and A. V. Proeyen, ``Fayet-Iliopoulos terms in supergravity without gauged R-symmetry'', JHEP \textbf{04}, 032 (2018) [arXiv:1712.08601 [hep-th]]

\bibitem{jpd} 
H.~Jang and M.~Porrati,
``Inflation, gravity mediated supersymmetry breaking, and de Sitter vacua in supergravity with a K\"ahler-invariant Fayet-Iliopoulos term,''
Phys. Rev. D \textbf{103}, no.10, 105006 (2021)
doi:10.1103/PhysRevD.103.105006
[arXiv:2102.11358 [hep-th]]. 

% KKLT mechanism
\bibitem{KKLT} S. Kachru, R. Kallosh, A. D. Linde and S. P. Trivedi, `` De Sitter vacua in string theory,'' Phys.
Rev. D \textbf{68}, 046005 (2003) [hep-th/0301240]; S.~Kachru, R.~Kallosh, A.~D.~Linde, J.~M.~Maldacena, L.~P.~McAllister and S.~P.~Trivedi, ``Towards inflation in string theory,'' JCAP \textbf{10}, 013 (2003) [arXiv: hep-th/0308055 [hep-th]].

\bibitem{ACK}
Y.~Aldabergenov, A.~Chatrabhuti and S.~V.~Ketov,
``Generalized dilaton-axion models of inflation, de Sitter vacua and spontaneous SUSY breaking in supergravity,''
Eur. Phys. J. C \textbf{79}, no.8, 713 (2019)
[arXiv:1907.10373 [hep-th]].

\bibitem{bf}
P.~Breitenlohner and D.~Z.~Freedman,
``Positive Energy in anti-De Sitter Backgrounds and Gauged Extended Supergravity,''
Phys. Lett. B \textbf{115}, 197-201 (1982).


\bibitem{HJ_Dr_thesis}
H.~Jang,
``Locally Supersymmetric Effective Field Theories of Inflation,''
[arXiv:2206.13736 [hep-th]].

% Superconformal Tensor Calculus
\bibitem{STC} E.~Cremmer, B.~Julia, J.~Scherk, S.~Ferrara, L.~Girardello and P.~van Nieuwenhuizen,
``Spontaneous Symmetry Breaking and Higgs Effect in Supergravity Without Cosmological Constant,''
Nucl. Phys. B \textbf{147}, 105 (1979); E.~Cremmer, S.~Ferrara, L.~Girardello and A.~Van Proeyen,
``Yang-Mills Theories with Local Supersymmetry: Lagrangian  , Transformation Laws and SuperHiggs Effect,''
Nucl. Phys. B \textbf{212}, 413 (1983); S. Ferrara, R. Kallosh, A. V. Proyen, and T. Wrase, ``Linear versus non-linear supersymmetry, in general,'' JHEP \textbf{04}, 065 (2016) [arXiv:1603.02653 [hep-th]].
% Soft SUSY breaking examples
\bibitem{fvp}
D.~Z.~Freedman and A.~Van Proeyen,
``Supergravity,''   Cambridge University Press (2012).

\bibitem{jp2}
H.~Jang and M.~Porrati,
``Component actions of liberated $ \mathcal{N} $ = 1 supergravity and new Fayet-Iliopoulos terms in superconformal tensor calculus,''
JHEP \textbf{11}, 075 (2021)
doi:10.1007/JHEP11(2021)075
[arXiv:2108.04469 [hep-th]].

\bibitem{GSR}
A.~Djouadi \textit{et al.} [MSSM Working Group],
``The Minimal supersymmetric standard model: Group summary report,'' [arXiv:hep-ph/9901246 [hep-ph]].

% High SUSY breaking in hidden sector
\bibitem{HighSUSY} S. A. R. Ellis and J. D. Wells, ``High-scale Supersymmetry, the Higgs Mass and Gauge Unication,'' Phys. Rev. D \textbf{96}, 055024 (2017) [arXiv:1706.00013v2 [hep-ph]]; E. Dudas, T. Gherghetta, Y. Mambrini, and K. A. Olive, ``Inflation and high-scale supersymmetry with an EeV gravitino,'' Phys. Rev. D \textbf{96}, 115032 (2017) [arXiv:1710.07341 [hep-ph]].

\bibitem{WeinbergSUSY}
S.~Weinberg, ``The quantum theory of fields. Vol. 3: Supersymmetry,'' Cambridge University Press (2013).

% LightScalar
\bibitem{LightScalar}
V.~Vennin, K.~Koyama and D.~Wands, ``Inflation with an extra light scalar field after Planck,''
JCAP \textbf{03}, 024 (2016) [arXiv:1512.03403 [astro-ph.CO]].


% EeV-scale Gravitino as a DM candidate
\bibitem{EeVGravitino} Emilian Dudas, Yann Mambrini, and Keith A. Olive, ``Case for an EeV Gravitino,'' Phys. Rev. Lett. \textbf{119}, 051801 (2017) [arXiv:1704.03008 [hep-ph]].

\bibitem{HeavyGravitino} K. A. Meissner and H. Nicolai, ``Standard Model Fermions and Infinite-Dimensional R-Symmetries,'' Phys. Rev. Lett. \textbf{121}, 091601 (2018) [arXiv:1804.09606 [hep-th]].

\bibitem{GDM} K. A. Meissner and H. Nicolai, ``Planck mass charged gravitino dark matter,'' Phys. Rev. D \textbf{100}, 035001 (2019) [arXiv:1809.01441 [hep-ph]].




\bibitem{Fermion_mass_matrix}
B.~Fuks, M.~Klasen, S.~Schmiemann and M.~Sunder, ``Realistic simplified gaugino-higgsino models in the MSSM,'' Eur. Phys. J. C \textbf{78}, 209 (2018) [arXiv:1710.09941 [hep-ph]].




\bibitem{WIMPZILLA1}
H.~Ziaeepour, ``A Decaying ultra heavy dark matter (WIMPZILLA): Review of recent progress,''
Grav. Cosmol. Suppl. \textbf{6}, 128-133 (2000)
[arXiv:astro-ph/0005299 [astro-ph]];


\bibitem{WIMPZILLA2}
J.~C.~Park and S.~C.~Park,
%``A testable scenario of WIMPZILLA with Dark Radiation,''
Phys. Lett. B \textbf{728}, 41-44 (2014)
doi:10.1016/j.physletb.2013.11.027
[arXiv:1305.5013 [hep-ph]].

\bibitem{WIMPZILLA3}
A.~Farzinnia and S.~Kouwn, ``Classically scale invariant inflation, supermassive WIMPs, and adimensional gravity,''
Phys. Rev. D \textbf{93}, no.6, 063528 (2016) [arXiv:1512.05890 [hep-ph]].

\bibitem{WIMPZILLA4}
E.~W.~Kolb and A.~J.~Long, ``Superheavy dark matter through Higgs portal operators,'' Phys. Rev. D \textbf{96}, no.10, 103540 (2017) [arXiv:1708.04293 [astro-ph.CO]].

\bibitem{SuperheavyDM}
E.~Alcantara, L.~A.~Anchordoqui and J.~F.~Soriano, ``Hunting for superheavy dark matter with the highest-energy cosmic rays,''
Phys. Rev. D \textbf{99}, no.10, 103016 (2019) [arXiv:1903.05429 [hep-ph]].





\end{thebibliography}
\end{document}